\documentclass{article}

\pdfoutput=1
 \usepackage{arxiv}

\usepackage[utf8]{inputenc} 
\usepackage[T1]{fontenc}    

\usepackage[square,sort,comma,numbers]{natbib}
\usepackage{lineno,hyperref}
\usepackage[title]{appendix}
\usepackage{amsmath}
\usepackage{array}
\usepackage{amssymb}
\usepackage{bm}
\usepackage{algorithm}
\usepackage{algpseudocode}
\usepackage{enumerate}
\usepackage{subcaption}
\usepackage{paralist}
\usepackage{multirow}
\usepackage{color}
\usepackage{todonotes}
\usepackage{xspace}
\usepackage{doi}
\usepackage{bigdelim,booktabs,colortbl}
\usepackage{nicefrac}

\newcommand{\openfoam}{\textsf{OpenFOAM}\xspace}
\newcommand{\comsol}{\textsf{COMSOL}\xspace}
\newcommand{\vulkan}{\textsf{VULCAN-CFD}\xspace}

\makeatletter
\newcommand{\xRightarrow}[2][]{\ext@arrow 0359\Rightarrowfill@{#1}{#2}}
\makeatother

\title{Inverse asymptotic treatment: \\ capturing discontinuities in fluid flows via equation modification}

\author{ \hspace{1mm}Shahab Mirjalili\thanks{Corresponding author}\ \ \thanks{Both authors contributed equally to this work.} \\
	Center for Turbulence Research, Stanford University, Stanford, CA 94305, USA\\
	\texttt{ssmirjal@stanford.edu} \\
	\And
	\hspace{1mm} S{\o}ren Taverniers\footnotemark[2] \\
	Palo Alto Research Center (PARC), 3333 Coyote Hill Road, Palo Alto, CA 94304, USA\\
  \And
	\hspace{1mm} Henry Collis \\
	Center for Turbulence Research, Stanford University, Stanford, CA 94305, USA\\
  \And
	\hspace{1mm} Morad Behandish \\
	Palo Alto Research Center (PARC), 3333 Coyote Hill Road, Palo Alto, CA 94304, USA\\
  \And
	\hspace{1mm} Ali Mani \\
	Center for Turbulence Research, Stanford University, Stanford, CA 94305, USA\\
  \texttt{alimani@stanford.edu} \\
}

\renewcommand{\shorttitle}{\textit{arXiv} Template}

\begin{document}
\maketitle

\begin{abstract}

  A major challenge in developing accurate and robust numerical solutions to multi-physics problems is to correctly model evolving discontinuities in field quantities. These manifest themselves as interfaces between different phases in multi-phase flows, or as shock and contact discontinuities in (either single- or multi-phase) compressible flows. A plethora of bespoke discretization schemes have been developed to capture both types of discontinuities in physics-based simulations. However, when a quick response is required to rapidly emerging challenges such as the need to design novel hypersonic vehicles, the complexity of such implementations impedes a swift transition from problem formulation to computation. This is exacerbated by the need to compose multiple interacting physics, which may cause specialized schemes to break unexpectedly as governing equations need to be changed and/or added. We introduce ``inverse asymptotic treatment'' (IAT) as a unified framework for capturing discontinuities in fluid flows that enables building {\it directly computable models} based on standard, off-the-shelf numerics. By capturing discontinuities through modifications at the level of the governing equations, rather than via reliance on specialized discretization schemes, IAT can seamlessly handle additional physics and thus makes it easier for novice end users to quickly obtain numerical results for a variety of multi-physics scenarios. This strategy also facilitates the use of a ``multi-physics'' compiler that automates the conversion of the modified PDEs to numerical source code readable by popular computational frameworks like \openfoam. We outline IAT in the context of phase-field modeling of two-phase incompressible flows, and then demonstrate its {\it generality} by showing how localized artificial diffusivity (LAD) methods for single-phase compressible flows can be viewed as instances of IAT. Through the real-world example of a laminar hypersonic compression corner, we illustrate IAT's ability to, within a span of just a few months, generate a directly computable model whose wall metrics predictions for sufficiently small corner angles come close to that of NASA's state-of-the-art \vulkan solver. Finally, we propose a novel LAD approach via ``reverse-engineered'' PDE modifications, inspired by total variation diminishing (TVD) flux limiters, to eliminate the problem-dependent parameter tuning that plagues traditional LAD. Through canonical numerical tests, we demonstrate that this ``limiter-inspired'' LAD approach, when combined with second-order central differencing, can robustly and accurately model compressible flows.

\end{abstract}


\section{Motivation and background} \label{sec:motivation}

Many fluid flow problems involve abrupt transitions in various quantities that must be viewed as evolving/moving discontinuities in the continuum assumption. The first category of discontinuities, {\it phase interfaces}, includes subcritical capillary interfaces that separate two immiscible phases. In subcritical conditions where pressure is not very high, such phase interfaces may be considered as discontinuities since their physical thickness is on the order of $1 ~\text{nm}$, which is much smaller than the molecular mean free path. The need for modeling phase interfaces as part of the simulation of multi-phase flows has spurred the development of numerous numerical discretization schemes that can broadly be divided into: (a) methods which move/deform the mesh to conform it to the discontinuities as they change over time; and (b) algorithms that track or capture the discontinuities within a non-conforming background mesh. In the context of modeling phase interfaces in fluid dynamics simulations, schemes with non-conforming meshes are preferred \citep{Tryggvason2011,Mirjalili_ARB}. Specifically, various one-fluid models that either track or capture the interface within a non-conforming mesh have been found to be more suitable than two-fluid models that conform the mesh with the interfaces \citep{Mirjalili_ARB}. The reason is that, generally, conforming-mesh methods require frequent re-meshing due to the possibility of mesh distortion in fluid dynamics, resulting in inaccurate and inefficient computations. These issues are exacerbated in two-phase flows because of frequent topological changes\citep{Tryggvason2011,Mirjalili_ARB}. Among one-fluid models, interface-\textit{capturing} approaches that belong to the classes of volume of fluid (VoF) \citep{Hirt1981,Xiao2005,Tryggvason2011}, level-set \citep{Sethian1999,Tryggvason2011,Olsson2005}, and phase-field (diffuse interface) \citep{Anderson1998,Saurel2018,Mirjalili_boundedness} have been found to be more advantageous than interface-\textit{tracking} methods \citep{Tryggvason2001,Tryggvason2011} which explicitly track the interface. This advantage is mainly due to the interface-capturing schemes' ability to automatically handle topological changes, such as in break-up or coalescence in two-phase flows. Moreover, explicit interface-tracking often results in scalability issues \textcolor{olive}{\citep{Xu2005,Ahmad2018}}.

The second category of fluid-flow discontinuities involves {\it shock and contact discontinuities} in transonic and supersonic flows, which also have a thickness on the order of the molecular mean free path and equally need to be viewed as discontinuities in a continuum formulation. The development of specialized numerical methods for their modeling has largely mirrored the strategy described above for multi-phase flows. Because of the high cost of frequent re-meshing, conforming-mesh strategies for compressible flows such as \cite{Montojo2017} are less popular than methods involving non-conforming grids that either locate and track shock and contact discontinuities developing as part of the system dynamics using shock-fitting \citep{Moretti1974,Marsilio1989}, or capture them via either complex high-resolution schemes or artificial diffusivity methods. High-resolution shock-capturing algorithms include the Monotonic Upstream-centered Scheme for Conservation Laws (MUSCL) \citep{vanLeer1979}, the essentially non-oscillatory (ENO) method \citep{Harten1987}, and its improved variants such as the weighted essentially non-oscillatory (WENO) \citep{Liu1994} and the targeted essentially non-oscillatory (TENO) \citep{Fu2016} schemes. Localized artificial diffusivity methods include the works by Cook and Cabot \citep{Cook2005} and Mani et al. \citep{Mani2009}.

Discontinuity-capturing schemes for both phase interface and shock/contact discontinuity modeling follow one of two distinct strategies. In the first approach, the task of capturing discontinuities is fulfilled by the \textit{numerical algorithm}---often the spatial discretization scheme---which adapts in the vicinity of discontinuities.
Examples of this strategy include VoF methods for capturing phase interfaces in two-phase flows, and MUSCL (through the use of a flux limiter) and ENO/WENO/TENO schemes (via a dynamic set of stencils) for capturing shock and contact discontinuities in compressible flows. Despite the successful adoption of such algorithms in a wide range of applications, they suffer from two major downsides:
\begin{enumerate}
    \item \textit{Achieving a discontinuity-capturing spatial discretization scheme that is both computationally efficient and load-balanced is still an active area of research.} On one hand, in interface-capturing methods like VoF \citep{Mirjalili_comparison,Jofre2015} and shock-capturing schemes like ENO and hybrid WENO schemes \citep{Zhao2019,Pirozzoli2002}, computational cost is heavily concentrated in regions with discontinuities; however, this makes parallel load-balancing complicated. On the other hand, the computations in methods like WENO and TENO are the same for all grid points in the domain, making load-balancing straightforward. Nevertheless, due to the computation of smoothness indicators and the required characteristic decompositions, these computations can be much more expensive than that of non-dissipative numerical schemes such as central differencing.
\item \textit{When adding new physics components to a problem, these highly-specialized schemes may break down in unexpected ways, and making ``surgical'' code modifications to resolve such issues is a difficult task for non-expert end users.} For instance, the original ENO/WENO schemes suffer from smearing at contact discontinuities. This has resulted in the development of techniques for sharpening the contact discontinuities \citep{Harten1987,Shu1989,Jiang1996}. Similarly, for compressible multi-phase flows, it is common to use interface-sharpening techniques to overcome the phase-interface smearing caused by upwind-biased schemes such as MUSCL and WENO \citep{Tiwari2013,Schmidmayer2020}.
\end{enumerate}

In contrast, the second category of discontinuity-capturing schemes are PDE-based methods that treat the discontinuities \textit{at the level of the governing PDEs}. In such strategies, capturing discontinuities is done by {\it designing artificial terms} that augment the original PDEs, rather than through the spatial discretization. This makes it possible to use standard \footnote{By ``standard'' we refer to methods that are commonly available to users in off-the-shelf commercial and open-source computational frameworks like \comsol \cite{Comsol2022} or \openfoam \cite{Jasak2007}. In this work, we refer more specifically to the use of central differencing.} numerical methods throughout the computational domain. This approach, in turn, enables the use of automation tools like a ``multi-physics compiler'' to convert the modified PDEs into numerical source code that is directly computable, i.e., readable by popular computational frameworks like \openfoam. Examples of such PDE-based strategies include phase-field models for capturing phase interfaces \citep{Mirjalili_boundedness,Jain2020,Khanwale2020}, and localized artificial diffusivity (LAD) schemes for capturing shock and contact discontinuities \citep{Cook2005,Kawai2010}.

Recent research has shown that there can be benefits to \textit{combining the above two discontinuity-capturing strategies}. Namely, in the context of compressible two-phase flows, \cite{Collis2022} showed that addition of regularizing terms on the PDE level (which are instances of IAT in action) can improve the robustness and accuracy of predictions using WENO and TENO spatial discretization. A discussion of such hybrid approaches is beyond the scope of this work.

\subsection{Outline}
\label{subsec:outline}

We introduce ``inverse asymptotic treatment'' (IAT) as a unifying recipe for systematically deriving PDE modifications that enables using standard spatial schemes effectively and uniformly throughout the domain, including in regions with discontinuities. We motivate and explain the concept of IAT, and list a set of requirements for a PDE modification strategy to be considered an IAT method, in Section \ref{sec:reverse_asymptotics}. In Section \ref{sec:inverse_asymptotics_2phase}, we illustrate these concepts for the case of two-phase fluid flows, highlighting IAT's advantages, including its ability to systematically enforce consistency. The rest of the article is devoted to applying IAT to capture discontinuities in compressible flows. In Section \ref{sec:compressible_flows}, we first demonstrate that traditional LAD methods can be considered as instances of IAT. Next, to circumvent problem-dependent tuning in traditional LAD methods, we present a novel method based on IAT that takes inspiration from flux-limiter-based numerical schemes with the total variation diminishing (TVD) property \citep{Harten1983} to ``reverse engineer'' a so-called ``limiter-inspired'' LAD method. Through a suite of numerical results for canonical shock problems and a real-world laminar hypersonic compression corner, we demonstrate the challenges with traditional LAD schemes, and how the limiter-inspired approach can resolve some of these issues. In Section \ref{sec:dir_comput_models}, using a set of key metrics, we evaluate the ability of IAT to generate directly computable models. In Section \ref{sec:PDE_mod_strategies}, we then present some general guidelines for PDE modification strategies beyond just IAT. Finally, conclusions and further extensions of the work are reserved for Section \ref{sec:conclusions}.

\section{Inverse asymptotic treatment} \label{sec:reverse_asymptotics}
\subsection{Motivation}
IAT is inspired by, and can be viewed conceptually as an inversion of, the method of matched asymptotic expansions (MAE) \citep{Verhulst2005,Lagerstrom2013} used extensively in fluid dynamics. MAE is commonly used to solve singularly perturbed PDEs by finding multiple approximate solutions, each of which is valid in a different region of the computational domain. The governing equations of the sub-problems (i.e., inner and outer solutions) are asymptotic limits of a more generally applicable equation (e.g., the Navier--Stokes equations). At the inter-region boundaries, the solutions are asymptotically matched as part of the MAE process.

The idea behind IAT is to treat sharp discontinuities (e.g., phase interfaces and shock waves) in a way that resembles inverting the MAE process, i.e., given physical discontinuities, we modify the original set of governing PDEs such that we can discretize them with a standard numerical method everywhere in the domain, including around the discontinuities, without incurring significant diffusion or dispersion errors. Below, we present an intuitive description of the theoretical bases of MAE and IAT.

Many problems in fluid dynamics (and other branches of physics) involve multiple scales. After appropriate scaling and non-dimensionalization, the governing equations for these problems can be considered as perturbations to reduced equations \footnote{For example, the Navier--Stokes equations can be seen as a perturbation to the Stokes equations.} in terms of a smallness parameter. Some problems are \emph{regular} perturbation problems, whereby the solution to the problem with the smallness parameter is slightly perturbed with respect to the solution without the smallness parameter. For instance, consider a steady creeping flow where the Reynolds number $Re\ll1$ is the smallness parameter. In this case, the spatial inertial term in the Navier--Stokes equations has a factor of $Re$ and can be safely ignored. In contrast, for \emph{singular} perturbation problems, the solution to the general problem (i.e., including the smallness parameter) does not converge to that of the limit problem (i.e.,  without the smallness parameter) as the smallness parameter tends to zero. Singular perturbation problems often arise when the smallness parameter is multiplied by the highest-order derivative in space (or time). The solutions to such problems typically involve small length (or time) scales, e.g., in the form of a thin layer (or short transience). For instance, consider a high-$Re$ flow, where after non-dimensionalization, $1/Re$ is the smallness parameter multiplying the viscous terms which contain the highest derivative in space. In this case, ignoring the viscous term would result in invalid solutions that do not satisfy the no-slip boundary conditions.

Singular perturbation problems are solved analytically or semi-analytically using MAE \citep{Verhulst2005,Leal2007,Lagerstrom2013}. As a concrete example, consider the flow around an airfoil with a moderately high $Re$ (i.e., it has not yet transitioned to turbulent conditions), shown schematically in the top panel of Figure \ref{fig:asympt_schematics}. The solution to this problem can be decomposed into an inner solution, valid in a thin ``boundary layer'' near the airfoil surface, and an outer solution, which holds outside of the boundary layer. The flow near the airfoil surface is dominated by viscosity and can be solved using boundary layer equations and the application of a no-slip boundary condition, while the flow outside this region can be treated as a potential flow, ignoring the no-slip boundary conditions and the thin boundary layer to which the viscous effects are confined. The result is a discontinuity in the velocity field near the airfoil surface (slip used to obtain the outer solution versus no-slip used to obtain the inner solution). The inner and outer solutions are coupled by matching the asymptotic velocity in the boundary layer, which is assumed to be infinitely thin when viewed from the outer region. \footnote{This approach is particularly useful for analytical treatments \citep{Verhulst2005,Lagerstrom2013}. More recently, high-performance computing (HPC) has enabled solving the Navier--Stokes equations for such problems in a more straightforward manner, resolving the boundary layer with a denser mesh near the boundary, without introducing a discontinuity or multiple regions governed by different equations.}

IAT can be seen as the \textit{inverse of MAE}, i.e., we start from a physically sharp transition (such as a phase interface or shock wave) and artifically thicken it by modifying the governing equations, as opposed to starting from a physically finite transition zone (the boundary layer inner zone) and artificially reducing it to an infinitely thin region when looking at it from the outer zone. The IAT approach results in artificially generated thickened transition zones that can be resolved on a computational mesh with parallelizable computation using standard numerical discretization schemes. The original PDE with an inherent solution discontinuity is replaced with a modified PDE whose solution exhibits a transitional inner zone and an outer zone in space, all of which are resolved automatically by numerical simulation using a standard numerical scheme everywhere in the domain.
The inner zone in this approach need not necessarily be governed by equations grounded in physics, but always involves an artificial smallness parameter $\epsilon \ll 1$ that governs its thickness (non-dimensionalized with a characteristic length). Often times, $\epsilon$ is controlled by the mesh size such that the discontinuity becomes resolvable over a few grid elements.

\subsection{Requirements for IAT}
\label{sec:IAT_reqs}
To be considered an IAT strategy, a PDE modification method needs to satisfy the following conditions:
\begin{enumerate}
    \item Modifications to the governing PDEs must not invalidate fundamental principles such as conservation laws or the laws of thermodynamics.

    \item The inner solution must hold in a thin zone whose thickness $\delta_\text{in}$ diminishes as the artificial smallness parameter is reduced, i.e., $\delta_\text{in}\propto\epsilon^{\alpha_1}$ with $\alpha_1>0$.

    \item The time scale to equilibrium $\tau_\text{in}$ for the inner solution must be smaller than all of the other relevant time scales of the problem, and must decrease as the smallness parameter decreases, i.e., $\tau_\text{in}\propto\epsilon^{\alpha_2}$ with $\alpha_2>0$. In other words, the inner zone must be governed by a quasi-steady solution that is controlled by the outer solution.

     \item Modifications to the governing PDEs should not introduce additional requirements in terms of spatial and temporal resolution. In other words, the thickness of the inner zone should be resolvable with the mesh, and the timescale of inner zone equilibration should not impose additional restrictions on the time step size.

    \item The outer solution must be consistent with the expected solutions away from the inner zone. For instance, an artificial diffusivity term in compressible flows should vanish away from shock / contact discontinuities such that the original equations are recovered in the outer regions. Moreover, the speed of the discontinuities and jump conditions across them should not be affected by the PDE modifications.
    
    \item The solution to the full inner-outer system of equations must converge to the discontinuous solution of the original (unmodified) equations in the ``sharp-interface limit'' defined by $\epsilon \rightarrow 0^+$.

\end{enumerate}
Such conditions must be verified by the physicist who designs the modifications to the equations.

\section{An inverse asymptotic treatment for two-phase flows} \label{sec:inverse_asymptotics_2phase}

\begin{figure}[ht!]
    \centering
    \includegraphics[width=\textwidth]{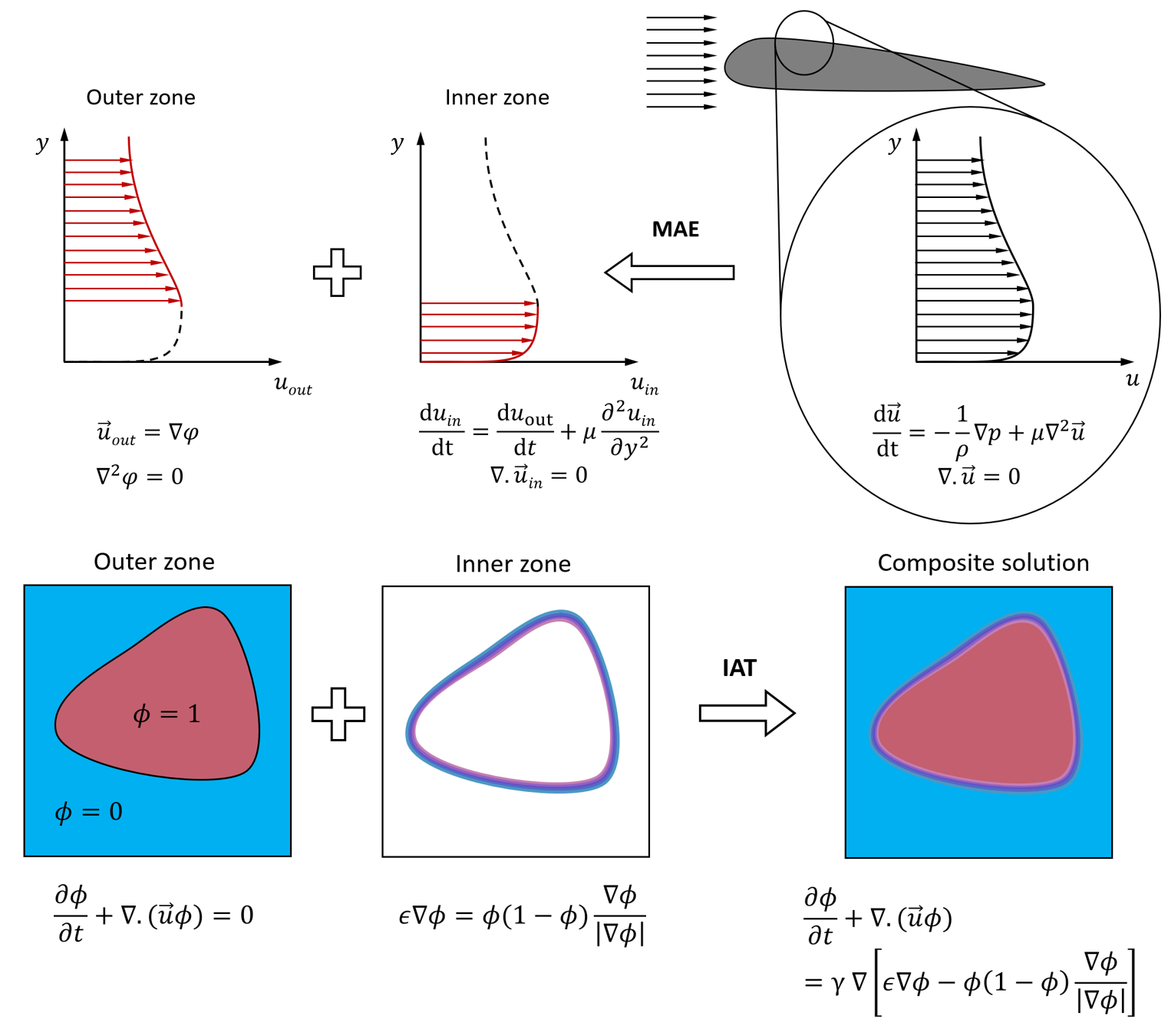}
    \caption{Top: schematic representation of the method of matched asymptotics, applied to solving a moderately-high $Re$ (i.e., still laminar) flow around an airfoil. Bottom: schematic representation of the inverse asymptotic treatment for modeling two-phase flows with the second-order conservative phase-field method.} \label{fig:asympt_schematics}
\end{figure}

To show how IAT works, let us first consider flows involving phase interfaces. A schematic representation of IAT for modeling incompressible, immiscible two-phase flows is presented in the bottom panel of Figure \ref{fig:asympt_schematics}. The original problem involving a sharp discontinuity has a governing PDE given
by $\partial \phi/\partial t + \nabla \cdot (\vec{u}\phi) = 0$ and is shown in the left-most sub-panel. Here, $\phi$ is the phase indicator function, which is 1 in one phase and 0 in the other phase. The governing equation states that the material derivative of the phase indicator variable is zero for incompressible, immiscible two-phase flows without phase change. In the outer zone, consisting of the regions outside of the discontinuity where $\phi$ is either $1$ or $0$, this equation is trivially satisfied ($0+0=0$). However, it is ill-defined across the discontinuity (interface) and must be interpreted in a weak sense where one must seek a viscous solution satisfying the entropy conditions \citep{Sethian1999}. Instead, we can augment the phase equation with two new terms, called the diffusion and sharpening terms, whose balance leads to the formation of a thin inner zone, as shown in the middle sub-panel of Figure \ref{fig:asympt_schematics}. This allows us to obtain a modified PDE whose solution is composed of an inner zone part and outer zone part (composite solution in the right-most sub-panel).

\subsection{Modified PDE and verification of IAT requirements}
\label{subsec:modifiedPDE_verification_IAT_req}

The composite equation used in the IAT described above is the second-order conservative phase-field model of \citep{Chiu_and_Lin,Mirjalili_boundedness}:
\begin{equation}
    \frac{\partial\phi}{\partial t}+\nabla\cdot(\vec{u}\phi)=\nabla\cdot\gamma \Big(\epsilon\nabla\phi-\phi(1-\phi)\vec{n} \Big), \label{eqn:2phase_pf}
\end{equation}
where $\vec{n} := \nabla\phi/|\nabla\phi|$ is the interface normal vector. All length and time scales are nondimensionalized by the outer length and advection time scale for the outer zone problem. Notice that the pair of artificial terms added to the right-hand side are conservative due to their factorization into a flux divergence. This property is needed to keep the conservation of volume (and mass) of each phase intact after the modification in an incompressible system, in line with the first requirement stated earlier.

Two dimensionless free parameters, $\epsilon$ and $\gamma$, have been introduced by IAT here. The smallness parameter $\epsilon \ll 1$ is multiplied by the highest-order spatial derivative, converting the original PDE into a singular perturbation problem. This introduces a thin inner zone wherein the diffusive flux $\epsilon\nabla\phi$ balances the sharpening flux $-\phi(1-\phi)\nabla\phi/|\nabla\phi|$. This balance leads to an approximately hyperbolic tangent profile along the normal direction \citep{Chiu_and_Lin,Mirjalili_boundedness,Jain2020}, $\phi \approx 1+\tanh{s(\phi)/(2\epsilon)}$, where $s(\phi)$ is the signed distance function from the interface ($\phi=0.5$ contour) and the dimensionless inner zone thickness scales with the smallness parameter, i.e., $\delta_\text{in}\sim\epsilon$. As such, the inner zone in this problem is indeed thin and becomes thinner proportionally to the smallness parameter, which is in agreement with the second requirement stated earlier.
 To satisfy the third requirement, the time scale to attain balance between diffusion and sharpening must be much smaller than the advection time scale for the outer zone problem. This means that $\tau_\text{in}\sim\epsilon/\gamma\ll1$. As such, we need the second parameter $\gamma$, representing the {\it speed} of diffuse interface equilibration, to regulate the magnitude of the artificial terms and separate the time scales between the inner and outer solutions. It has to be large enough, such that $\tau_\text{in} \ll 1$, enabling neglection of the left-hand side (variation and advection) terms in Eq.~\eqref{eqn:2phase_pf} in the inner zone, leading to a quasi-steady equilibrium between the artificial diffusion and sharpening terms, whose solution in the $\epsilon \to 0^+$ limit converges to the hyperbolic tangent profile over an $\epsilon-$thin region. At the same time, if $\gamma$ is too large, we risk making $\tau_\text{in}\sim\epsilon/\gamma$ too small, which renders the IAT too stiff for numerical solution, noting that the inner zone time and length scales must be numerically resolved using explicit time-stepping schemes. This is an instance where the computational physicist designing the PDE modification must provide guidelines for the free parameters of the artificial terms. In \cite{Mirjalili_boundedness}, it was shown that if $\epsilon\sim\Delta x$, where $\Delta x$ represents the spatial grid size, choosing $\gamma\sim |\vec{u}|$ is sufficient to ensure robust and accurate solutions for Eq.~\eqref{eqn:2phase_pf} without incurring additional numerical stiffness, in line with the fourth requirement for IAT stated above.

Away from the interface, the solution to Eq.~\eqref{eqn:2phase_pf} in the outer zone is given by either $\phi=0$ or $\phi=1$, leading to zero diffusion ($\nabla \phi = 0$) and zero sharpening ($\phi(1-\phi) = 0$) which is consistent with the discontinuous solution to the original PDE prior to modifications, as depicted by the bottom-left sub-panel of Figure \ref{fig:asympt_schematics}. Moreover, as explained in \cite{Mirjalili_boundedness}, the interface velocity is not affected by the addition of the artificial terms and the solution of the modified PDE converges to the original PDE's discontinuous solution in the sharp-interface limit. Therefore, Eq.~\eqref{eqn:2phase_pf} satisfies the fifth and sixth requirements described earlier.

It is important to note that had we omitted either of the right-hand side terms, our PDE modification would not have been an instance of IAT. Specifically, without the diffusive flux, the PDE would not be a singular perturbation problem governed by a smallness parameter. Instead, the interface would remain discontinuous and the sharpening term would be identically zero everywhere, as if it was not included at all. On the other hand, omitting the sharpening term while keeping the diffusion term results in a singular perturbation problem. However, because there is no physical sharpening mechanism in the original PDE (in contrast to shock generation in compressible flows), the interface thickness grows in time incessantly, invalidating requirements 2 and 3. After a while, diffusion corrupts solutions away from the interface, violating requirement 5 as well. These two examples demonstrate that not all PDE modifications are useful or can be considered as instances of IAT.

\subsection{Systematic enforcement of consistency}
\label{subsec:systematic_consistency}

One of the main advantages of treating discontinuities by modifying the underlying PDE---as opposed to making changes to numerical schemes to handle discontinuities---is that physical principles and insight can be leveraged to consistently account for the effect of new terms in the coupled system of equations. For instance, it has been shown in the literature that a conservative mass flux must be propagated into the momentum transport equation to enforce thermodynamic consistency \citep{ Terashima2013,Haga2019,Jain2020, Huang2021,Mirjalili2021}. This is also necessary to satisfy the first requirement stated earlier. More specifically, the phase field equation (conservation of mass) is coupled with the equation expressing conservation of momentum:
\begin{equation}
    \frac { \partial (\rho \vec{ u } ) }{ \partial t } +\nabla \cdot \Big((\rho \vec{ u }-\vec{S} )\otimes \vec{ u } \Big) =-\nabla P+\nabla \cdot \Big(\mu (\nabla \vec{ u } +\nabla ^\mathrm{T}\vec{ u } ) \Big)+\vec{ F }_B,
    \label{eqn:2phase_momentum}
\end{equation}
where $P$ is the pressure field, $\rho$ and $\mu$ are the density and dynamic viscosity computed by linear interpolation of the phase field via

\begin{equation}
    \rho = \rho_1\phi+\rho_2(1-\phi) ,
    \label{eqn:2phase_density}
    \end{equation}
    where $\rho_1$ and $\rho_2$ are the densities of pure phase 1 ($\phi=1$) and pure phase 2 ($\phi=0$), respectively, and
    \begin{equation}
    \mu = \mu_1\phi+\mu_2(1-\phi),
    \label{eqn:2phase_viscosity}
\end{equation}
where $\mu_1$ and $\mu_2$ are the dynamic viscosities of pure phase 1 and pure phase 2, respectively. $\vec{F}_B$ represents the body forces, including surface tension. The artificial term $\vec{S}$ is the mass-momentum consistency correction flux, given by:
\begin{equation}
    \vec{S} := (\rho_1-\rho_2) \vec{R}, \quad\text{where}~ \vec{R} := \gamma \Big( \epsilon\nabla\phi-\phi(1-\phi)\vec{n} \Big),
    \label{eqn:mass_flux}
\end{equation}
where $\vec{R}$ is the artificial term inside the divergence operator on the right-hand side of Eq.~\eqref{eqn:2phase_pf}. Intuitively, the reason to include this term in the momentum equation is that the artificial mass transport due to $\vec{R}$ leads to an artificial momentum transport. As demonstrated in \cite{Mirjalili2021,Huang2021}, such a consistent modification of the PDEs accounting for the mass and momentum conservation laws automatically leads to conservation of kinetic energy remaining intact in spite of the modifications. This is significant, as naive modification of governing equations can have unintended consequences such as violating secondary conservation principles that are not explicitly enforced, leading to nonphysical numerical effects \cite{Perot2011}. Making consistent modifications that preserve such principles is much harder and less controllable at a numerical level when dealing with discretized equations or computational geometric manipulations.

The system of coupled equations given by Eq.~\eqref{eqn:2phase_pf} through Eq.~\eqref{eqn:mass_flux} satisfies all of the requirements listed earlier for IAT, and by virtue of thickening the interface in a controlled manner, can be used in conjunction with standard numerical schemes (e.g., central differencing)
to develop robust numerical solvers for practical two-phase flows, including those with high-density ratios and/or high $Re$ numbers \citep{Mirjalili2021}. In the case of phase field modeling via Eq.~\eqref{eqn:2phase_pf}, the form of the artificial terms was inspired by conservative level-set methods \citep{Olsson2005} and formally derived from the Allen--Cahn equation by \cite{Chiu_and_Lin}. Guidelines for the free parameters $\epsilon$ and $\gamma$ were provided by \cite{Mirjalili_boundedness} to ensure numerical stability and boundedness. This is not the only way to develop phase field models with such properties. Examples of other choices for the phase field equation that have been used to model two-phase flows are the Cahn--Hilliard equation \citep{Abels2012,Khanwale2020,Huang2021} and conservative Allen--Cahn equation \citep{Kim2014,Zhai2015,Jeong2017,Joshi2018,Huang2020}. These phase-field models can all be considered instances of IAT as well.
 The main difference between these options for IAT in two-phase flows is mostly related to the form of the RHS terms and the free parameters' values in the phase-field equation, since Eqs.~\eqref{eqn:2phase_momentum}--\eqref{eqn:mass_flux} can be derived based on thermodynamic principles, including conservation of mass, momentum and kinetic energy. Crucially, as long as the RHS terms in the phase-field equation are in conservative form and satisfy all the IAT requirements, coupling with Eqs.~\eqref{eqn:2phase_momentum}--\eqref{eqn:mass_flux} should result in a system of equations that can be used for modeling realistic two-phase flows.

\subsection{Application of IAT to scalar transport for two-phase flows}
\label{subsec:application_IAT_2phase}

A second, slightly more complicated, example pertains to modeling passive scalar (e.g., heat, charge, or chemical species) transport in two-phase flows. In the physical discontinuous setting, the equation governing the concentration of the species in each phase is an advection-diffusion equation,
\begin{equation}
    \frac{\partial {c}_i}{\partial t} + \nabla\cdot(\vec{u}{c}_i)=\nabla\cdot(D_i{\nabla}{c}_i),
    \label{eqn:mass_local}
\end{equation}
for $i=0,1$, where $D_i$ is the diffusivity and $c_i$ is the species' concentration, respectively, in phase $i$. Let us now consider a dilute species that is confined to one phase as it is insoluble in the other, such as salt in a two-phase water-air system. Now, one may be tempted to model the scalar concentration in the whole domain using a single scalar value $c$ and, in a similar fashion to Eq.~\eqref{eqn:2phase_density} and Eq.~\eqref{eqn:2phase_viscosity}, naively compute the diffusivity field as:
\begin{equation}
    D = D_1\phi+D_2(1-\phi).
    \label{eqn:2phase_naive_diffusivity}
\end{equation}
However, the coupled system in this case would result in artificial leakage of the species into the insoluble phase \citep{Jain2019,Mirjalili2022}. In other words, the expected outer solutions are not obtained and the fifth requirement of IAT is not satisfied. In \cite{Jain2019}, an alternative PDE for confined scalar transport consistent with the phase-field model was derived by making an analogy with the phase-field equation Eq.~\eqref{eqn:2phase_pf}. A more comprehensive IAT for scalar transport in arbitrary two-phase systems which also incorporated interfacial transfer was derived in \cite{Mirjalili2022}. There, a microstructure was assumed for the interface between the two phases, from which perturbation analysis, lubrication theory, and homogenization was used to derive the interfacial transfer terms and scalar transport model that can be used for arbitrary diffusivity ratios, including the confined scalar case. The derived scalar transport model in \cite{Mirjalili2022} satisfies all of the requirements for IAT listed above.

\section{Application of inverse asymptotics to compressible flows}
\label{sec:compressible_flows}

The IAT approach is applicable to a wide variety of fluid flows involving sharp discontinuities. We outlined IAT in Section \ref{sec:inverse_asymptotics_2phase} for two-phase flows, and numerical simulations using the resulting methods can be found in \cite{Mirjalili_boundedness,Mirjalili_comparison,Mirjalili2021,Jain2019,Mirjalili2022}. For the remainder of this work, we focus on single-phase compressible supersonic/hypersonic flows. Specifically, we present traditional LAD and a novel LAD approach inspired by total variation diminishing (TVD)-endowing flux limiters as examples of applying IAT to model single-phase compressible flows. In each section, we provide numerical results to illustrate the properties of each method.

The idea of augmenting physical viscosities with artificial viscosities for the purpose of numerically capturing shocks dates back to the pioneering work of \cite{vonneumann1950}. Cook and Cabot \cite{Cook2005} proposed a localized artificial viscosity method that can be used along with high-order compact differencing schemes to capture shock-turbulence interactions. Localized artificial viscosities are required to vanish in smooth well-resolved regions of the flow and provide dampening around shocks to capture discontinuities. Since then, the dynamic localized artificial diffusivity (LAD) approach---using high-order spatial derivatives to detect discontinuities and then filtering---has been widely adopted by other researchers. Specifically, \cite{Fiorina2007} extended the approach of \cite{Cook2005} to include artificial mass and species concentration diffusivities, allowing for capturing of contact and species discontinuities, respectively. Cook \cite{Cook2007} introduced artificial thermal conductivity into the energy equation for large-eddy simulations of compressible turbulent mixing, and Kawai and Lele \cite{Kawai2008} extended this formulation to curvilinear and anisotropic grids. Recognizing that the magnitudes of the strain-rate tensor and dilatation are similar near shocks while the latter is much smaller in turbulent vortices away from shocks, Mani et al. \cite{Mani2009} proposed a modification to the artificial bulk viscosity (ABV) model to limit the application of the ABV to shocks. In particular, they suggested localization based on Laplacians of the dilatation field and a shock ``sensor'' that would switch on in compression zones and switch off in expansion zones. A comparison study by Johnsen et al. \cite{Johnsen2010} showed that this modification results in significant improvement in predictions of compressible turbulence statistics compared with the original LAD approach of \cite{Cook2007}. Additionally, Kawai et al. \cite{Kawai2010} showed that adding a switch based on the Ducros-type sensor can allow for more finely tuned switch-off of the dilatation-based ABV in weakly compressible zones in addition to dilatational zones. More recently, Lee and Lele \cite{Lee2017} studied the performance of these LAD models in problems involving deflagration and detonation.

All of the above efforts utilized high-order (namely, sixth- to tenth-order) compact differencing schemes for spatial discretizations. The aforementioned advancements in localization strategies for LAD methods resulted in high-resolution, low-cost, easy-to-implement, and modular numerical methods for simulation of compressible turbulent flows involving shocks. Motivated by their advantages for simulating turbulent flows \citep{Moin2016} and their ease of implementation and adaptability, in this work we apply similar ideas using only second-order central difference (CD) spatial operations. We control the large dispersion errors of the second-order schemes by augmenting the mass, momentum, and total energy equations with artificial mass diffusivity, artificial shear/bulk viscosities, and artificial thermal conductivity, respectively. The governing equations, including the artificial diffusivities and the terms needed for thermodynamic consistency, are presented in Section \ref{subsec:trad_LAD}. Our numerical tests in Sections \ref{subsubsec:results_1D_trad_LAD} and \ref{subsubsec:results_2D_trad_LAD} show that adopting LAD schemes along with lower-order CD schemes, is not as successful as the high-order compact differencing schemes typically used in the literature. Specifically, we observe that when using CD for spatial discretization, problem-dependent tuning of the non-dimensional user-defined constants is required for acceptable levels of accuracy. Moreover, there are no guarantees of robustness for the LAD schemes. To circumvent these deficiencies, inspired by analysis of second-order total variation diminishing (TVD) flux limiters, in Section \ref{subsec:TVD_LAD} we present a novel approach for localizing the artificial diffusivity terms that, in contrast to previous phenomenological LAD schemes, achieves high accuracy and robustness using CD without problem-dependent tuning or de-aliasing of the solutions via filtering. We derive the model in a one-dimensional (1D) setting and assess its performance using canonical 1D tests in Section \ref{sec:results_tvd_inspired}. Finally, we present an approach for extending the proposed model to multi-dimensional settings using Cartesian grids in Section \ref{subsec:multiD_TVD_insp_LAD}.

\subsection{Traditional localized artificial diffusivity (LAD)}
\label{subsec:trad_LAD}

In order to regularize shock and contact discontinuities in compressible flows, one may take inspiration from physical diffusivities to introduce ``artificial diffusivities'' (ADs). Consistent modification of the compressible Navier--Stokes equations for a calorically perfect gas via ADs yields:
\begin{equation}
    \frac{\partial {\rho}}{\partial t} + \nabla\cdot({\rho}\vec{u})=\nabla\cdot\vec{m}^*,
    \label{eqn:mass_PDE}
\end{equation}
\begin{equation}
    \frac { \partial (\rho \vec{ u } ) }{ \partial t } +\nabla \cdot \left(\rho \vec{ u }\otimes \vec{ u } \right )+\nabla P=\nabla \cdot\left(\vec{m}^*\otimes \vec{ u } \right )+\nabla\cdot(\tau+\tau^*),
    \label{eqn:momentum_PDE}
\end{equation}
\begin{equation}
    \frac { \partial (\rho E  ) }{ \partial t } +\nabla \cdot \left[(\rho E+P)\vec{ u } \right ]=\nabla \cdot\left(\vec{m}^*\frac{|\vec{u}|^2}{2}\right )+\nabla\cdot\left[\vec{u}\cdot(\tau+\tau^*)\right]-\nabla\cdot(\vec{q}+\vec{q}^*).
    \label{eqn:energy_PDE}
\end{equation}
Here $\rho$ is the density, $\vec{u}$ is the velocity, $E=e+(|\vec{u}|^2/2)$ is the total energy per unit mass with $e$ the internal energy per unit mass, $P=(\gamma-1)\rho e$ is the pressure, and $\vec{q}=-k\nabla T$ is the heat flux with $k$ the thermal conductivity and $T=(\gamma-1)e/R$ the temperature, wherein $R$ is the gas constant and $\gamma$ is the polytropic coefficient of the gas. Furthermore, $\tau$ is the viscous stress tensor given by the constitutive relation
\begin{equation}
    \tau={\mu}[\nabla\vec{u}+(\nabla\vec{u})^T]+(\beta-2\mu/3)(\nabla\cdot\vec{u})\delta,
    \label{eqn:viscous_stress}
\end{equation}
where $\mu$ is the dynamic shear viscosity, $\beta$ is the bulk viscosity, and $\delta$ is the identity tensor.

The above equations are augmented explicitly with $\vec{m}^*$, $\tau^*$, and $\vec{q}^*$, which are the artificial mass flux, artificial viscous stress tensor, and artificial heat flux, respectively. The momentum and total energy equations are also augmented by further corrections; namely, $\nabla \cdot\left(\vec{m}^*\otimes \vec{ u } \right)$, $\nabla \cdot\left(\vec{m}^*\|\vec{u}|^2/{2}\right )$, and $\nabla\cdot\left(\vec{u}\cdot\tau^*\right)$, for thermodynamic consistency \citep{Jameson2008,Terashima2013,Haga2019}. The artificial viscous stress tensor $\tau^*$ and artificial heat flux $\vec{q}^*$ are computed analogously to their physical counterparts, replacing the physical diffusivities by the corresponding artificial ones (denoted by asterisks). The artificial mass flux $\vec{m}^*$ has no physical counterpart, and is computed via:
\begin{equation}
    \vec{m}^*=\chi^*\nabla\rho,
    \label{eqn:art_mass_flux}
\end{equation}
where $\chi^*$ is the artificial mass diffusivity (AMD).
For AMD, \cite{Fiorina2007} proposed the following formula:
\begin{equation}
      \chi^*={C}_{\chi}^{l}\frac{a_0}{c_P} \Delta^{2l+1}\overline{|\nabla^{2l}s|},
      \label{eqn:AMD_Fiorina}
\end{equation}
where the overbar is a truncated Gaussian filter \cite{Cook2007}, $a_0$ is the reference speed of sound, $c_P$ is the specific heat at constant pressure, $s$ is the entropy, $\Delta$ is a characteristic length based on the grid cell dimensions, and ${C}_{\chi}^{l}$ is an $O(1)$ user-defined constant with $l$ to be specified. The spatial derivatives of entropy are utilized so that the AMD activates around contact and material discontinuities, in addition to shocks.

Next, artificial shear viscosity is localized using:
\begin{equation}
    \mu^*={C}_{\mu}^{l}\overline{\rho \Delta^{2l+2}|\nabla^{2l}(\nabla\cdot\vec{u})H(-\nabla\cdot\vec{u})|},
    \label{eqn:ASV}
\end{equation}
where $\epsilon = \frac{1}{2} \Big(\nabla \vec{u} + (\nabla \vec{u})^\mathrm{T}\Big)$ is the strain-rate tensor, and ${C}_{\mu}^{l}$ is an $O(1)$ user-defined constant. Note that in Eq.~\eqref{eqn:ASV} we use spatial derivatives of the dilatation field rather than the magnitude of the strain rate tensor which was used in \cite{Cook2005}. This is to avoid numerical dissipation affecting vortical structures away from shocks and discontinuities, especially in turbulent regimes. For artificial bulk viscosity, based on \cite{Mani2009} we use:
\begin{equation}
    \beta^*={C}_{\beta}^{l}\overline{\rho \Delta^{2l+2}|\nabla^{2l}(\nabla\cdot\vec{u})H(-\nabla\cdot\vec{u})|},
    \label{eqn:ABV_Mani}
\end{equation}
while, following \cite{Cook2007}, artificial heat conductivity is localized via:
\begin{equation}
    k^*={C}_{k}^{l}\overline{\frac{\rho a_0}{T} \Delta^{2l+1}|\nabla^{2l}e|}.
      \label{eqn:AHC_Cook}
\end{equation}
In Eq.~\eqref{eqn:ABV_Mani} and Eq.~\eqref{eqn:AHC_Cook}, ${C}_{\beta}^{l}$ and $C_k^{l}$ are $O(1)$ user-defined constants.
\subsubsection{Verification of IAT requirements}
\label{subsubsec:verification_IAT_LAD}
We now demonstrate that traditional LAD methods satisfy all the requirements for IAT methods listed in Section \ref{sec:IAT_reqs}. Let us, for instance, consider the case of the momentum equation and artificial shear and bulk viscosities (Eqs.~\eqref{eqn:momentum_PDE}, \eqref{eqn:ASV}, and \eqref{eqn:ABV_Mani}).
\begin{enumerate}
    \item The conservative form of the artificial terms in Eq.~\eqref{eqn:momentum_PDE} automatically preserves conservation of momentum. Furthermore, the correction term involving $\vec{m}^*$ ensures consistency with the mass equation (Eq.~\eqref{eqn:mass_PDE}), and in turn is consistently accounted for in the energy equation (Eq.~\eqref{eqn:energy_PDE}) to guarantee thermodynamic consistency.
    \item By inspecting Eqs.~\eqref{eqn:ASV}, and \eqref{eqn:ABV_Mani}, it can be shown that the artificial viscosity values scale with $u\Delta$. Hence, the artificial smallness parameter is proportional to the mesh size $\Delta$. In the inner zone, the advective flux balances the artificial diffusive flux, from which we can show, via scaling arguments, that the inner zone thickness is $O(\Delta)$. Hence, the inner solution holds in a thin zone whose thickness diminishes as the artificial smallness parameter is reduced (i.e., as the mesh is refined).
    \item The time scale to equilibrium for the inner solution is given by balancing the temporal term and the diffusive term, yielding $\tau_{\text{in}}\sim\Delta/u$. This is the time scale for the propagation of information across a single cell. Even for coarse resolutions, this time scale is therefore expected to be much smaller than all the physical time scales of the problem. Additionally, this time scale decreases as the artificial smallness parameter is reduced since the latter is proportional to the mesh size.
    \item The thickness of the inner zone is on the order of the mesh size, and the time scale for the inner zone equilibrium is on the order of the time needed for information to propagate across a single cell, which dictates the CFL requirements for the simulation. Hence, the PDE modifications do not introduce more stringent spatial or temporal resolution requirements.
    \item By design, the artificial diffusivity terms vanish in smooth regions away from the discontinuities (i.e., the inner zone). As such, the original equations are recovered away from shock and contact discontinuities. Moreover, in the absence of physical viscosity, applying the Rankine-Hugoniot jump conditions in a control volume around a shock reveals that the shock speed is unaltered by the addition of localized artificial viscosities.
    \item As the mesh is refined, the artificial smallness parameter controlling the magnitude of the artificial viscosities and shocks tends to zero. Hence, in the absence of physical viscosity, the solution to the momentum equation (Eq. \eqref{eqn:momentum_PDE}) converges to the discontinuous (sharp) entropy solution \citep{Sethian1999} as the artificial smallness parameter is reduced to zero.
\end{enumerate}

\subsubsection{1D canonical problems}
\label{subsubsec:results_1D_trad_LAD}

Here we test the traditional LAD scheme on two canonical shocked flow problems: Sod's shock tube problem \cite{Sod1978} and the 1D Shu-Osher problem \cite{Shu1989}. Note that all numbers are dimensionless for the canonical problems in this work.

Sod's shock tube assesses the capability of a scheme in capturing moving shocks and contact discontinuities. On a unit-length spatial domain $[0, 1]$, at time $t = 0$, an initially stationary shock at $x = 0.5$ separates a region with (non-dimensional) density $\rho_l = 8$, speed $u_l = 0$, and pressure $P_l = 7.128$ on the left-end from a region with $\rho_r = 1$, $u_r = 0$ and $P_r = 0.712$ on the right-end, while $\gamma = 1.4$. A collocated uniform grid with $\Delta =0.005$ is employed and the time step is $\Delta t=0.0005$. The solution is assessed at time $t_f=0.2$.

The Shu-Osher problem provides a 1D idealization of shock-turbulence interaction and involves a Mach 3 shock propagating into a field with density perturbations. On a spatial domain $[-5, 5]$, at time $t = 0$, a shock at $x = -4$ separates a region with (non-dimensional) density $\rho_l = 3.857$, speed $u_l = 2.629$, and pressure $P_l = 10.333$ on the left-end from a region with $\rho_r = 1 + 0.2\sin(5x)$, $u_r = 0$ and $P_r = 1$ on the right-end, while $\gamma = 1.4$. A collocated uniform grid with $\Delta=0.025$ is employed and the time step is $\Delta t=0.001$. The solution is assessed at $t_f=1.8$. An important criterion for assessing the performance of a scheme on this problem is the accuracy of capturing the entropy waves in the post-shock zone. In particular, numerical dissipation from upwind-biased schemes dampens these physical oscillations. Similarly, it is desirable for an LAD scheme to be vanishing in this region with smooth fields.

\begin{figure}
\centering
      \begin{subfigure}{0.49 \linewidth}
\includegraphics[width=0.95\textwidth]{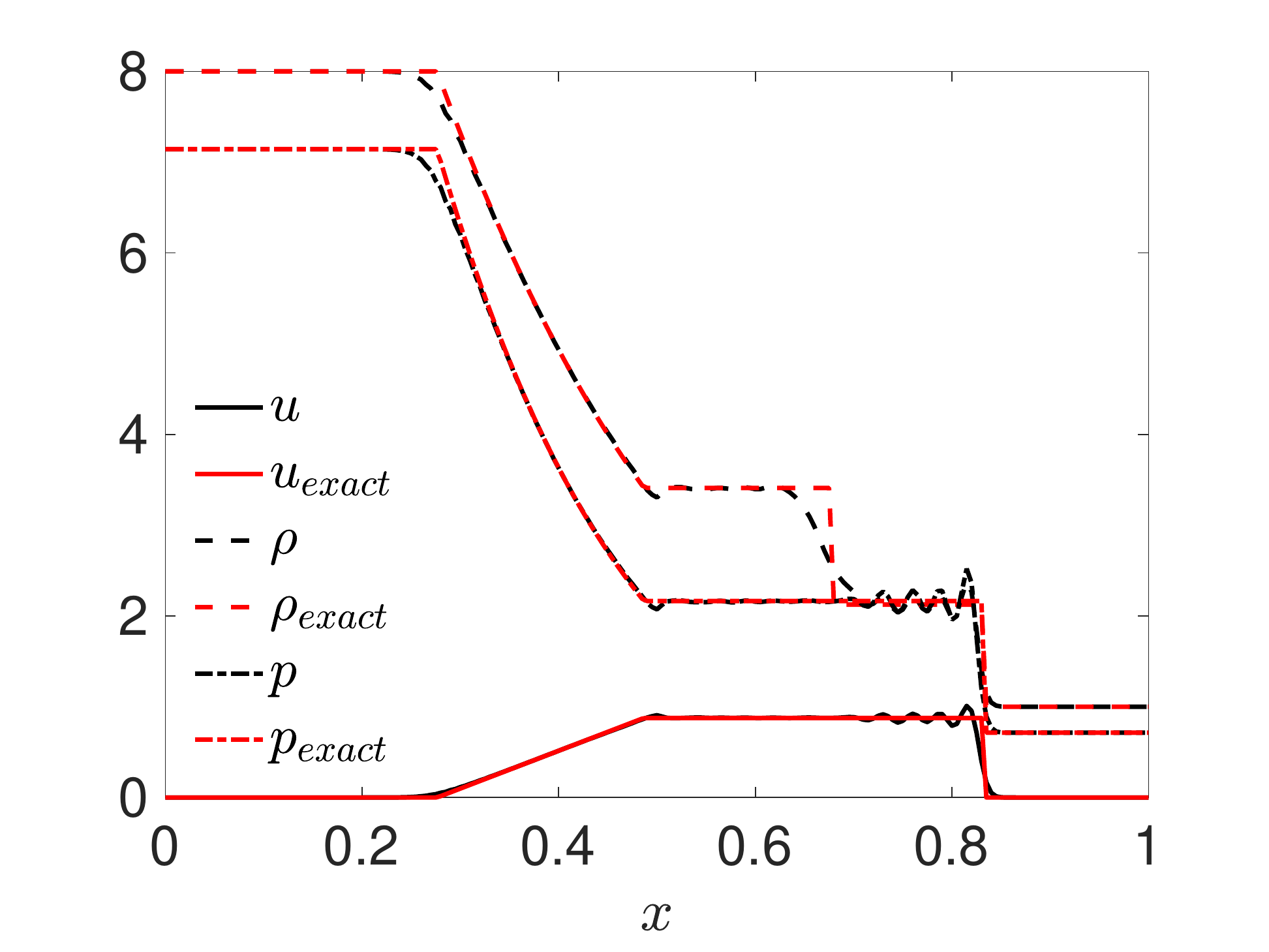}
  \caption{}
  \end{subfigure}
      \begin{subfigure}{0.49 \linewidth}
\includegraphics[width=0.95\textwidth]{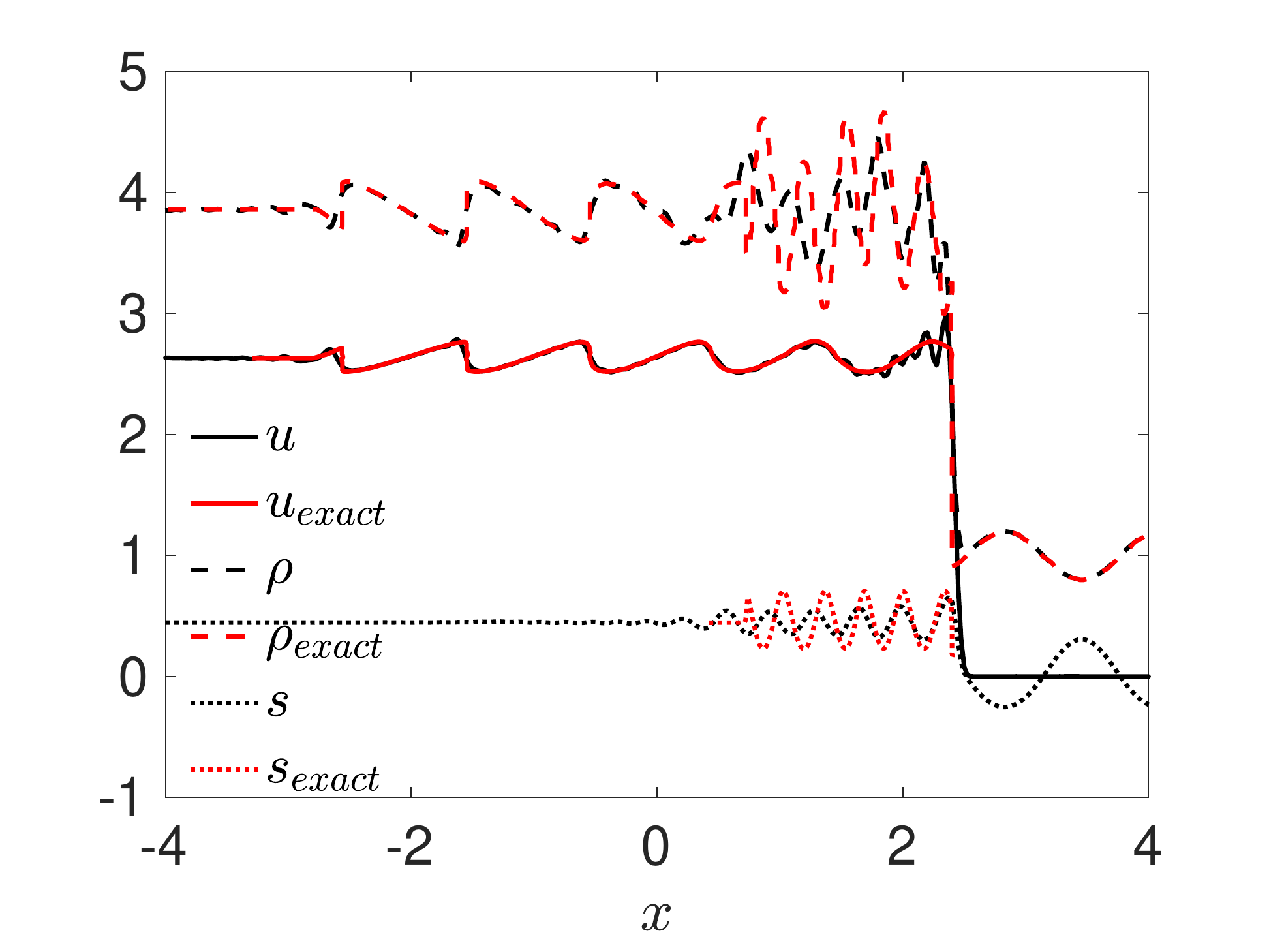}
  \caption{}
  \end{subfigure}
        \begin{subfigure}{0.49 \linewidth}
\includegraphics[width=0.95\textwidth]{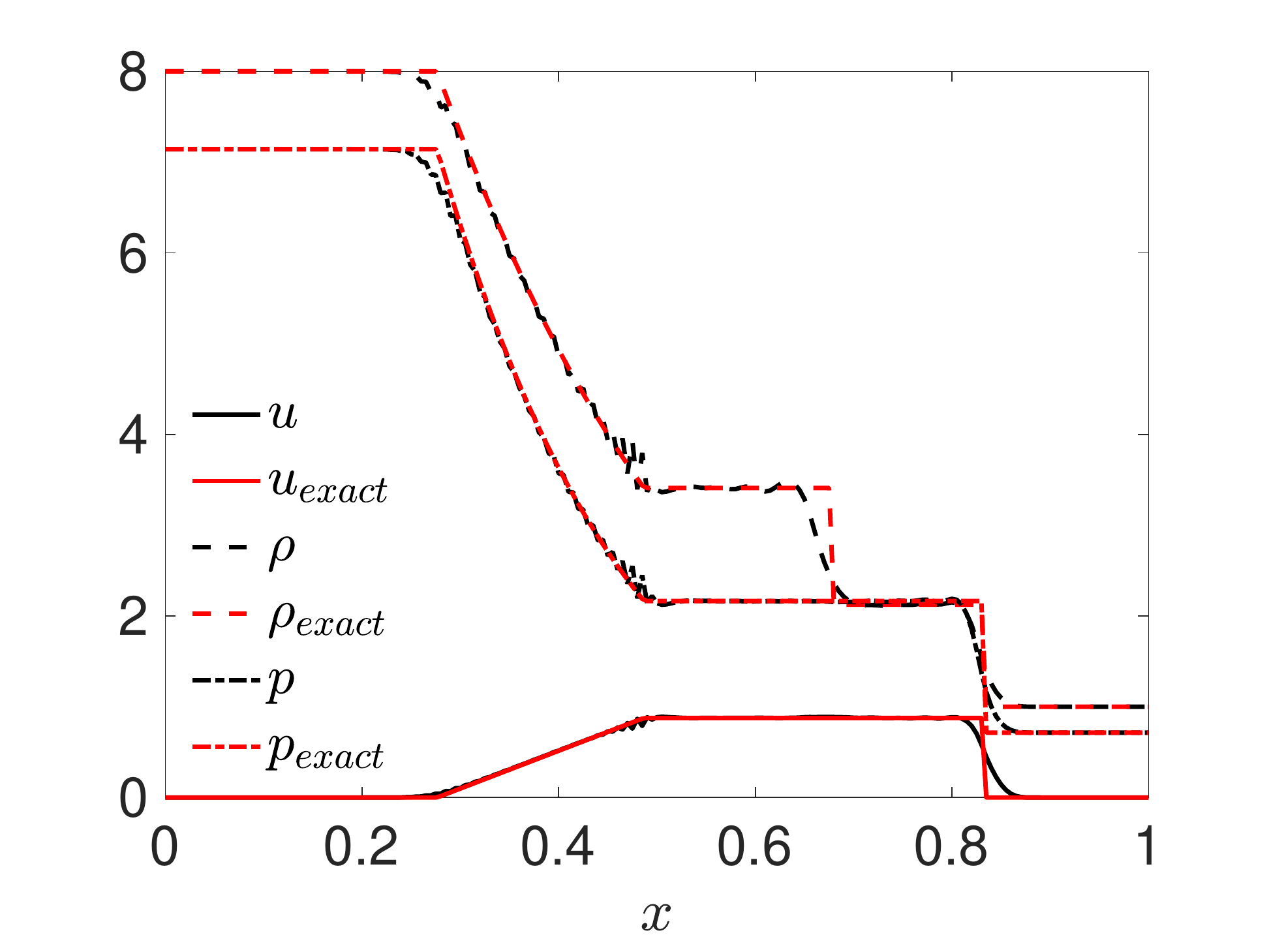}
  \caption{}
  \end{subfigure}
      \begin{subfigure}{0.49 \linewidth}
\includegraphics[width=0.95\textwidth]{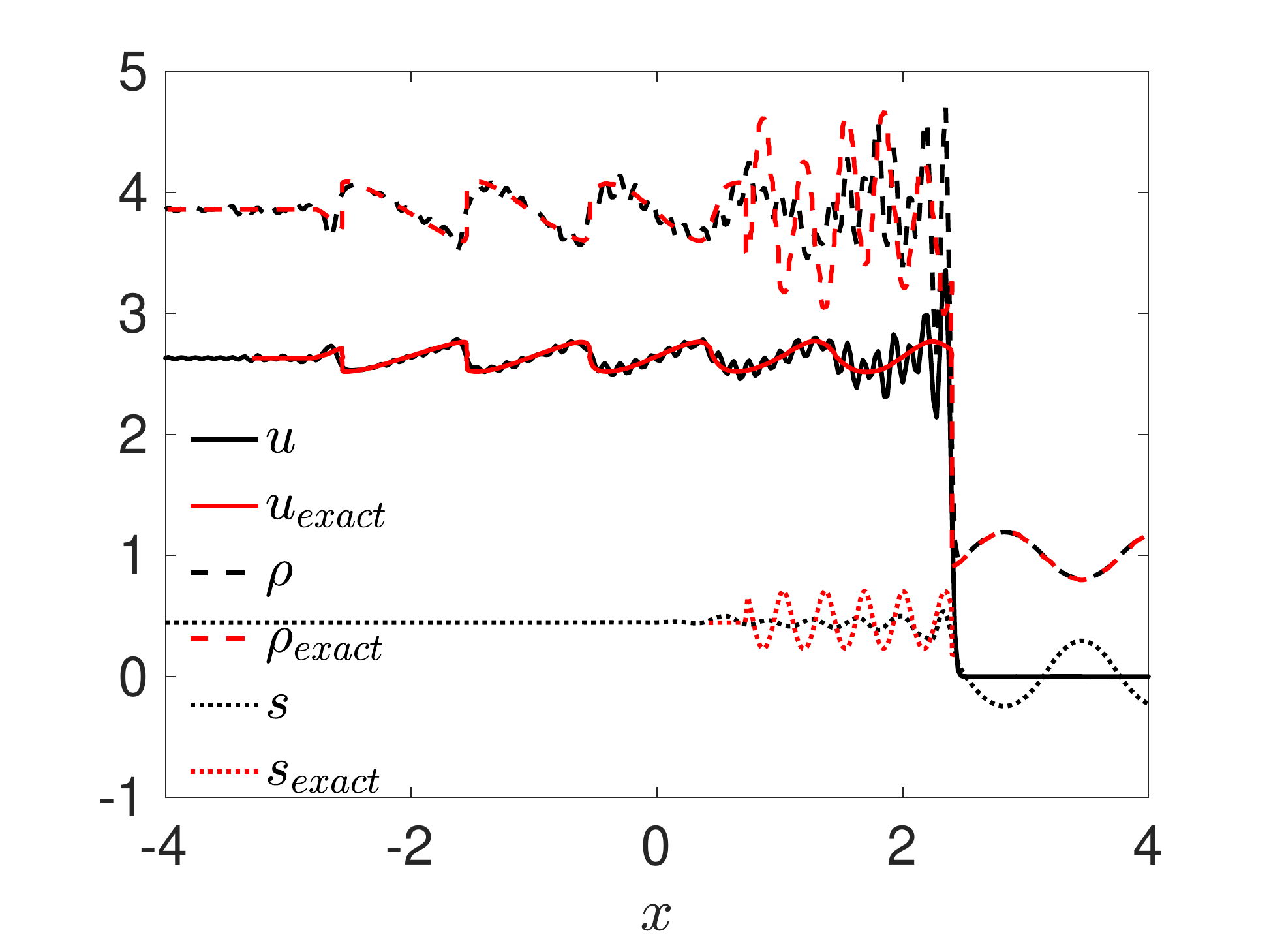}
  \caption{}
  \end{subfigure}
    \caption{ Simulation results using traditional LAD with optimal ${C}_{\chi}^{1}, {C}_{\beta}^{1/2}$, and ${C}_{k}^{1}$ values are plotted at the top panels for (a) Sod's shock tube and (b) Shu-Osher problems. Simulation results from traditional LAD when the parameter values in panels (a) and (b) are swapped are plotted on the bottom panels for (c) Sod's shock tube and (d) Shu-Osher problems. Notice that results degrade for both problems upon swapping the free parameter values, demonstrating significant problem-dependence for the CD+LAD approach.}
      \label{fig:LAD_optimal_stuff}
\end{figure}

We use traditional LAD with $l=1$ for mass and heat, and $l=1/2$ for momentum diffusion, in numerical experiments on the above problems, along with second-order CD for spatial discretization. In contrast to the boundedness analysis used in two-phase flows \citep{Mirjalili_boundedness}, a pen-and-paper analysis of the coupled system given by Eqs.~\eqref{eqn:mass_PDE}--\eqref{eqn:AHC_Cook} to guide the choice of the free parameters based on TVD property or positivity is not straightforward. Instead, we use the two described canonical 1D problems to optimize the free parameters of the LAD method. The effect of artificial shear and bulk viscosity is the same for 1D problems. As such, we perform a parameter sweep in the three-dimensional parameter space given by ${C}_{\chi}^{1}, {C}_{\beta}^{1/2}$, and ${C}_{k}^{1}$. Giving equal weight to capturing velocity, density, and pressure, the optimal set of parameters for Sod's shock tube and Shu-Osher problems are found separately. The simulation results for the two problems are shown in Figures \ref{fig:LAD_optimal_stuff}(a,b), using their respective optimal free parameter values. Despite optimization of the free parameters, we observe excessive oscillations at the discontinuities in Sod's shock tube problem (Figure \ref{fig:LAD_optimal_stuff}(a)). From Figure \ref{fig:LAD_optimal_stuff}(b), it is clear that using optimized free parameters for the Shu-Osher problem allows for the post-shock entropy waves and the physical velocity and density oscillations to be captured without excessive dampening. Crucially, as shown in Figure \ref{fig:LAD_optimal_stuff}(c-d), if we swap the optimal parameter sets for the two problems, we see significant degradation of the results for both problems, indicating how problem-dependent these parameters are. This suggests that the free parameters must be tuned separately for every new problem when using LADs in combination with CD. This is unappealing and practically impossible and the main reason we seek LAD schemes that do not require problem-dependent tuning in Section \ref{subsec:TVD_LAD}.

\begin{figure}
\centering
\begin{subfigure}{0.49 \linewidth}
\includegraphics[width=0.95\textwidth]{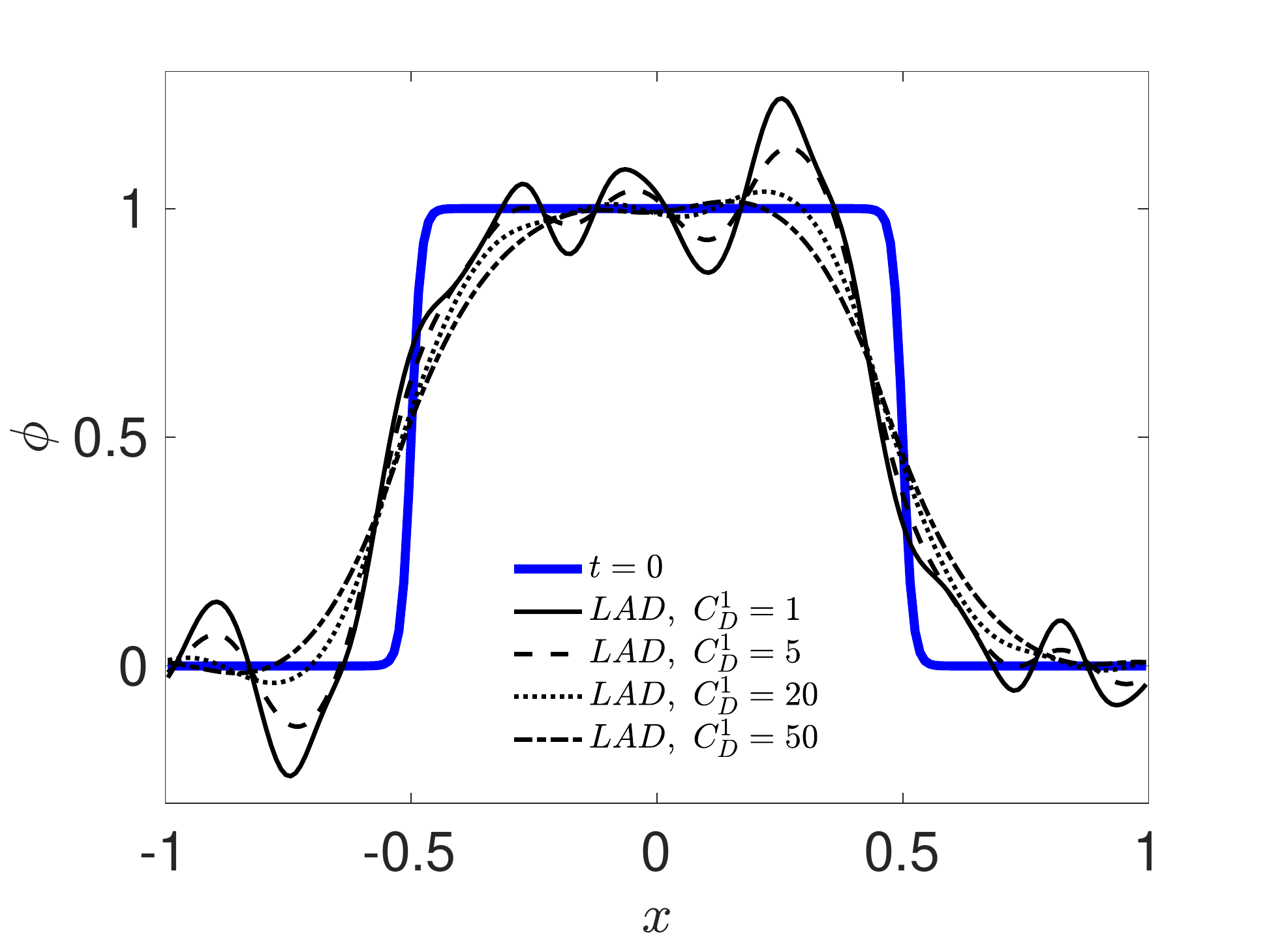}
  \caption{}
\end{subfigure}
\begin{subfigure}{0.49 \linewidth}
\includegraphics[width=0.95\textwidth]{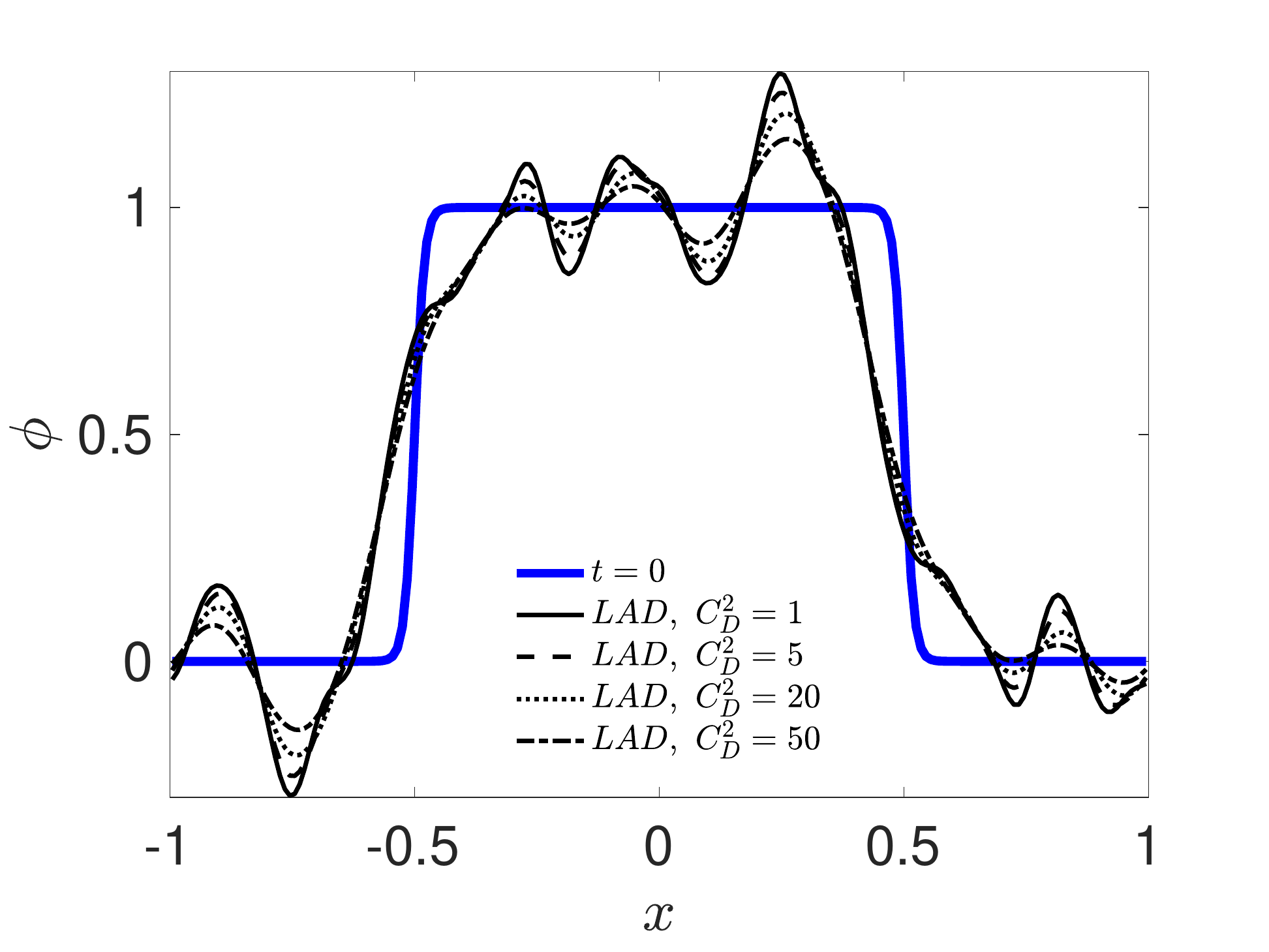}
  \caption{}
  \end{subfigure}
    \caption{Results from scalar advection Eq.~\eqref{eqn:PDE_LAD} using traditional LAD methods Eq.~\eqref{eqn:LAD_phi} plotted at $t_f=20$ for (a) $l=1$ and (b) $l=2$, demonstrating $\phi$ profiles that either suffer from excessive diffusion or dispersion errors.}
      \label{fig:1D_scalar_advection_LAD}
\end{figure}

To further illustrate the shortcomings of combining CD with traditional LAD schemes, let us consider a simple 1D problem where velocity is fixed to be $u=1$ to advect a scalar $\phi$ via:
\begin{equation}
    \frac{\partial\phi}{\partial t}+\frac{\partial(u\phi)}{\partial x}=\frac{\partial}{\partial x}(D^*\frac{\partial\phi}{\partial x}),
    \label{eqn:PDE_LAD}
\end{equation}
\begin{equation}
    D^*={C}_{D}^{l}|u|_\text{max}\Delta^{2l+1}\overline{|\nabla^{2l}\phi|},
    \label{eqn:LAD_phi}
\end{equation}
where $D^*$ is an artificial diffusivity and $C_D^l$ is a user-defined constant. Starting from a $\text{tanh}$-shaped initial profile, the drop is advected with RK4 time-stepping for 20 time units in a $[-1, 1]$ periodic domain, discretized with $\Delta=0.01$. Note that while $D^*$ is needed to stabilize the CD discretization, a global value would result in excessive diffusion. The simulation results using the localized model presented in Eq.~\eqref{eqn:LAD_phi} with $l=1,2$ are shown in Figure \ref{fig:1D_scalar_advection_LAD}, which exhibit excessive undershoots/overshoots as well as numerical diffusion of the solution. This is in accordance with the simulation results obtained for Sod's shock tube, where we observe either excessive oscillations or excessive diffusion at the discontinuities in Figure \ref{fig:LAD_optimal_stuff}(a, c).

\subsubsection{2D compression corner, laminar hypersonic flow}
\label{subsubsec:results_2D_trad_LAD}

\begin{figure}
   \centering
    \includegraphics[width=0.9\textwidth]{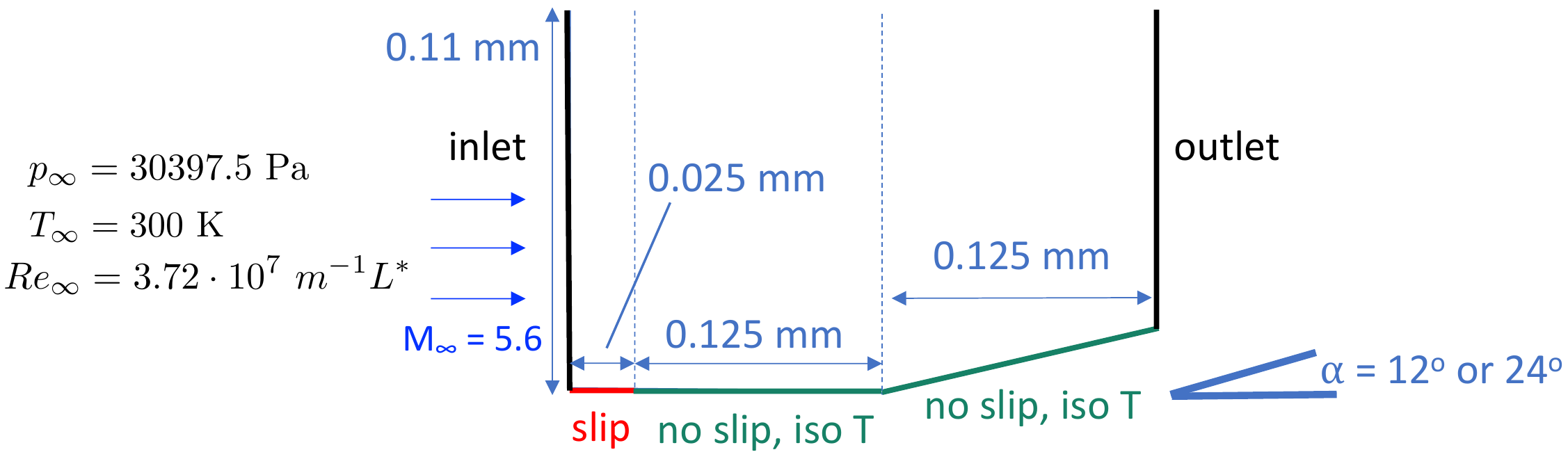}
    \caption{Schematic representation of the laminar compression corner used to study hypersonic SWBLI.}
    \label{fig:cc}
\end{figure}

Next, let us consider the more advanced problem of shock wave and boundary layer interaction (SWBLI) in a 2D laminar hypersonic flow past a compression corner. Accurate and robust modeling of supersonic (Mach number $\geq$ 1) and hypersonic (Mach number $\geq$ 5) flows is essential to the reliable operation and continued development of high-speed aircraft, as well as ensuring the safe atmospheric re-entry of manned spacecraft. A specific geometry of interest in these problems is the compression corner \cite{Anderson2020} (Figure \ref{fig:cc}), which leads to interactions between the shock waves (SW) and boundary layer (BL) that are formed near the surface of the flat plate and ramp section starting from the leading edge \cite{Gatski2013}. The SWBLI, especially in hypersonic compression corners with large turning angles, presents a number of computational challenges to accurately capture the physics of SW/SW and SW/BL interactions, laminar-to-turbulent transitions, and separation of the BL \citep{Smith1991,Korolev2002,Shvedchenko2009}. To make this problem more tractable for testing purposes, we consider a laminar setup based on the work in \cite{Shellabarger2018}. This setup is characterized by a hypersonic interaction parameter $\chi=M_{\infty}/Re_{\infty}=1.37$ that captures the strength of the SWBLI, with a free-stream Mach number of $M_{\infty}=5.6$ and a free-stream Reynolds number of $Re_{\infty}=3.72\times 10^7L^*\mathrm{m^{-1}}$, where $L^*=0.125\times 10^{-3}$ m is the length of the no-slip section of the flat plate (Figure \ref{fig:cc}).

\begin{figure}
 \centering
\begin{subfigure}{0.49 \textwidth}
\includegraphics[width=0.95\textwidth]{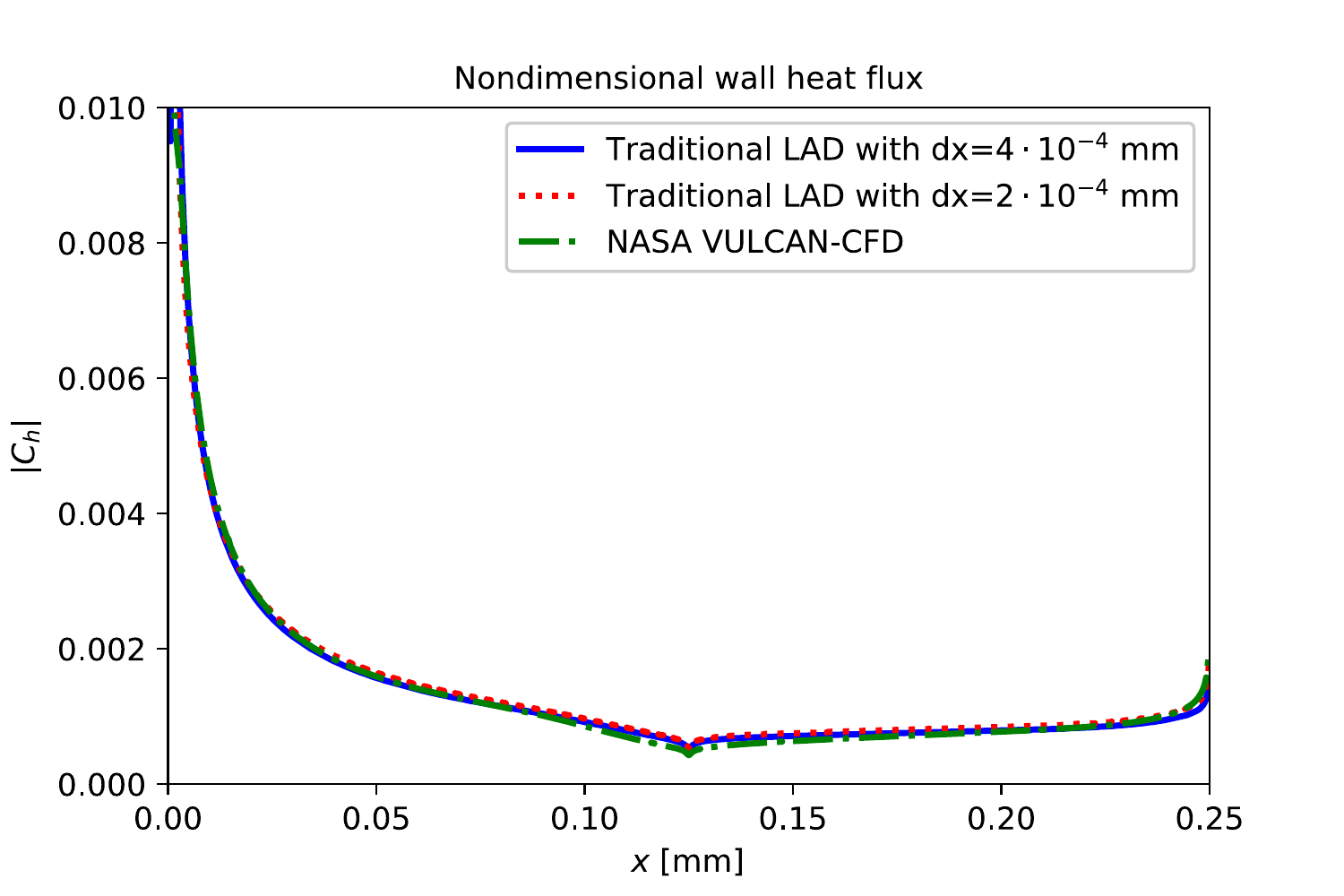}
  \caption{}
  \label{fig:C_h_12deg}
\end{subfigure}
\begin{subfigure}{0.49 \textwidth}
\includegraphics[width=0.95\textwidth]{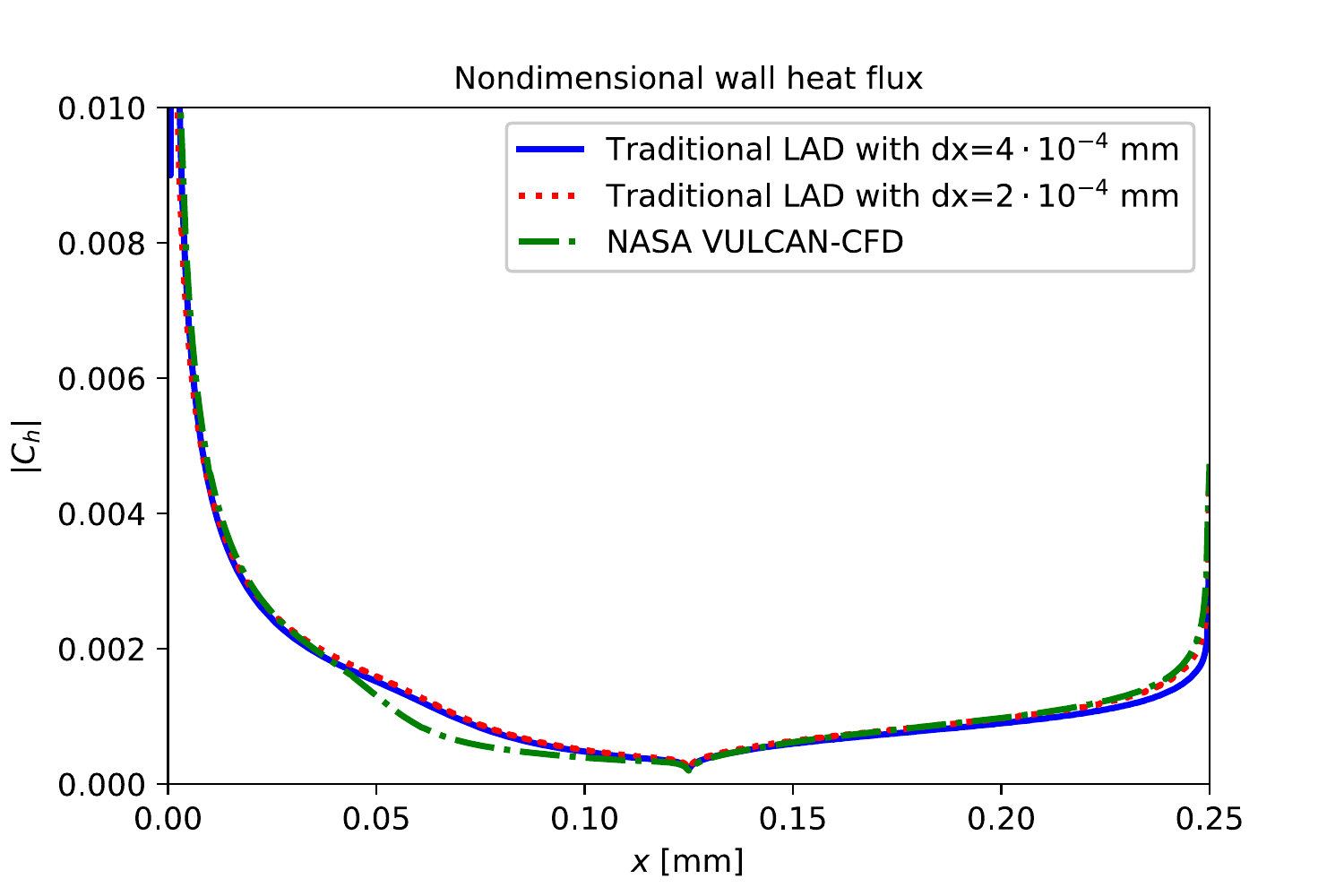}
  \caption{}
  \label{fig:C_h_24deg}
  \end{subfigure}
         \begin{subfigure}{0.49 \textwidth}
\includegraphics[width=0.95\textwidth]{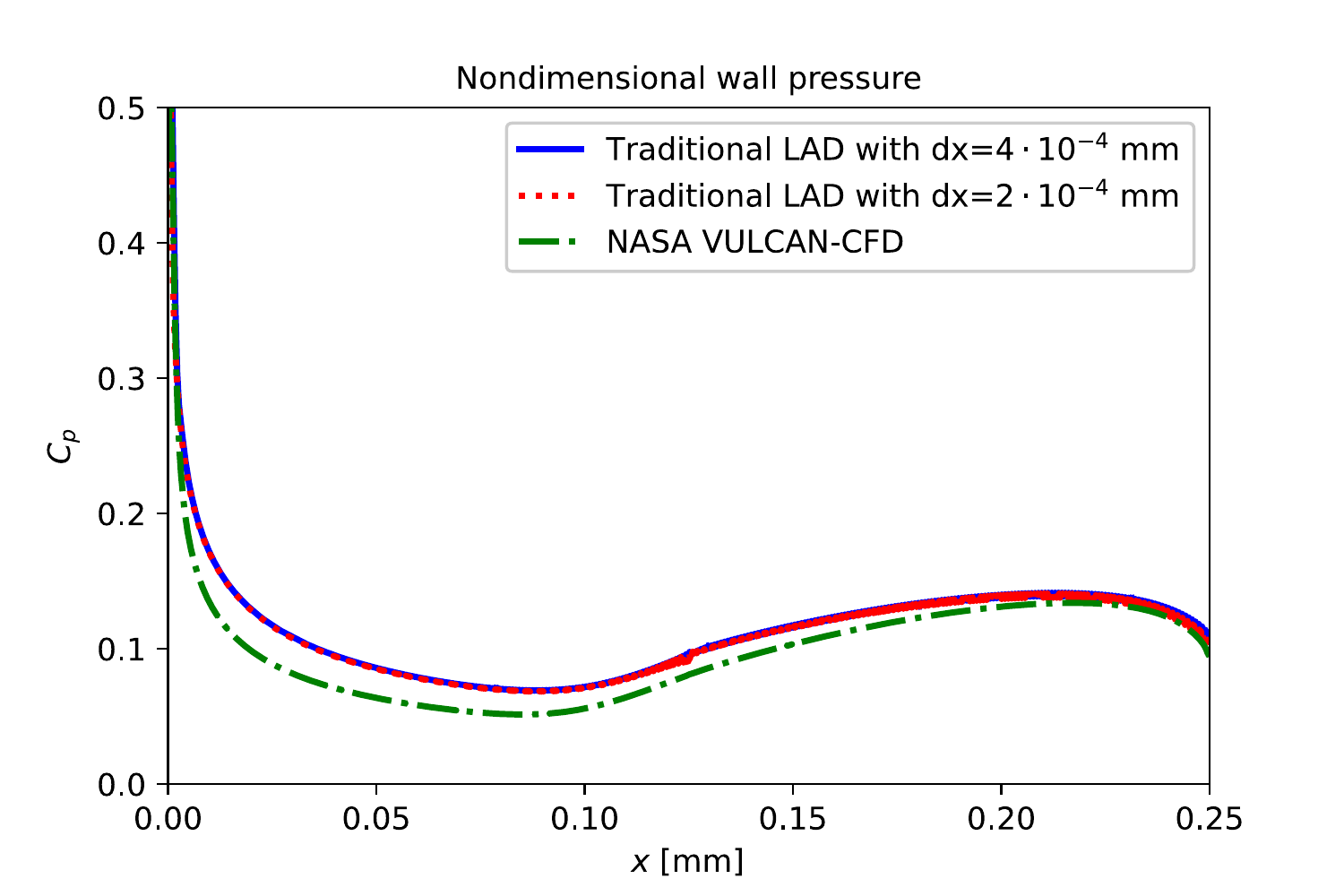}
  \caption{}
  \label{fig:C_p_12deg.eps}
  \end{subfigure}
         \begin{subfigure}{0.49 \textwidth}
\includegraphics[width=0.95\textwidth]{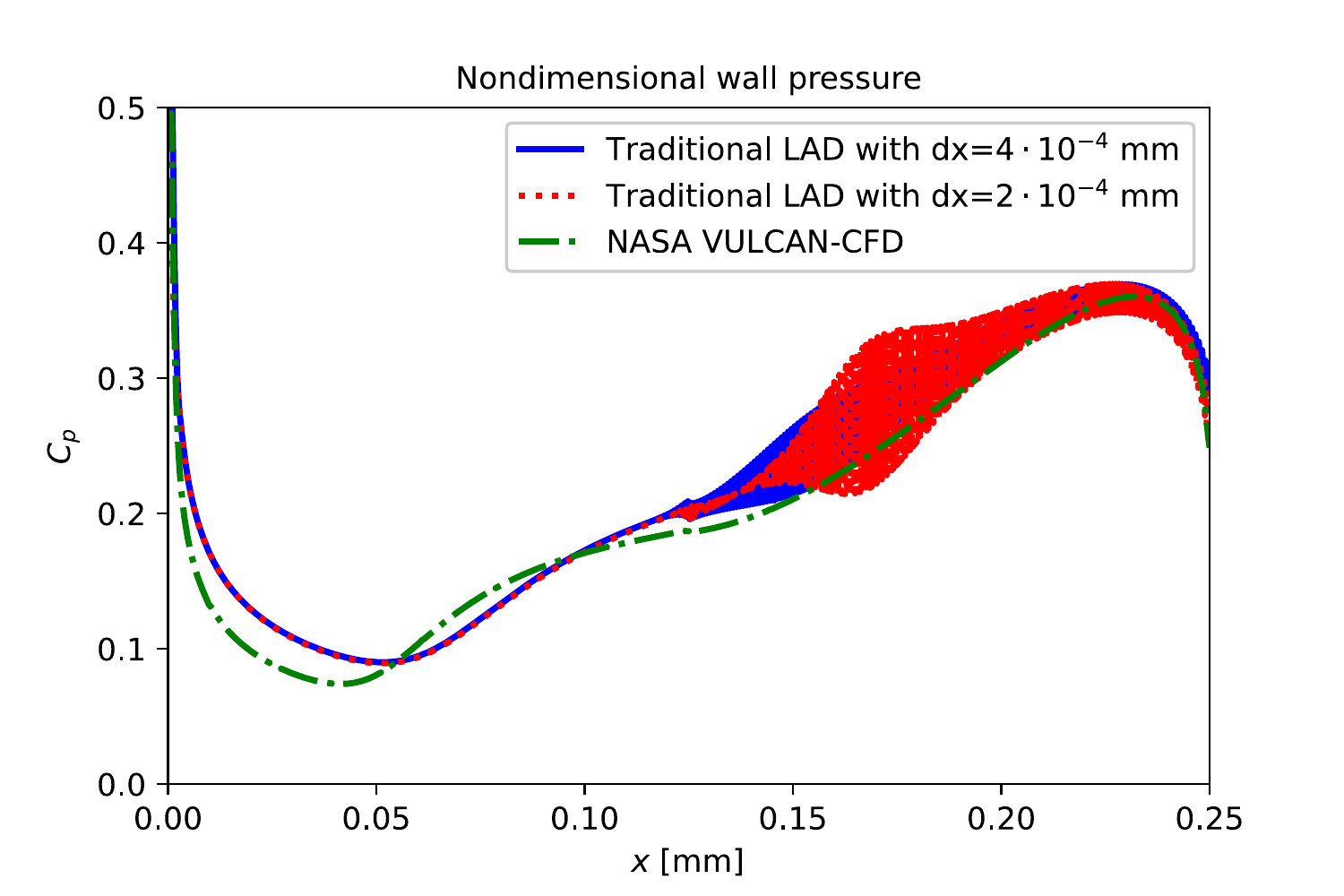}
  \caption{}
  \label{fig:C_p_24deg}
  \end{subfigure}
           \begin{subfigure}{0.49 \textwidth}
\includegraphics[width=0.95\textwidth]{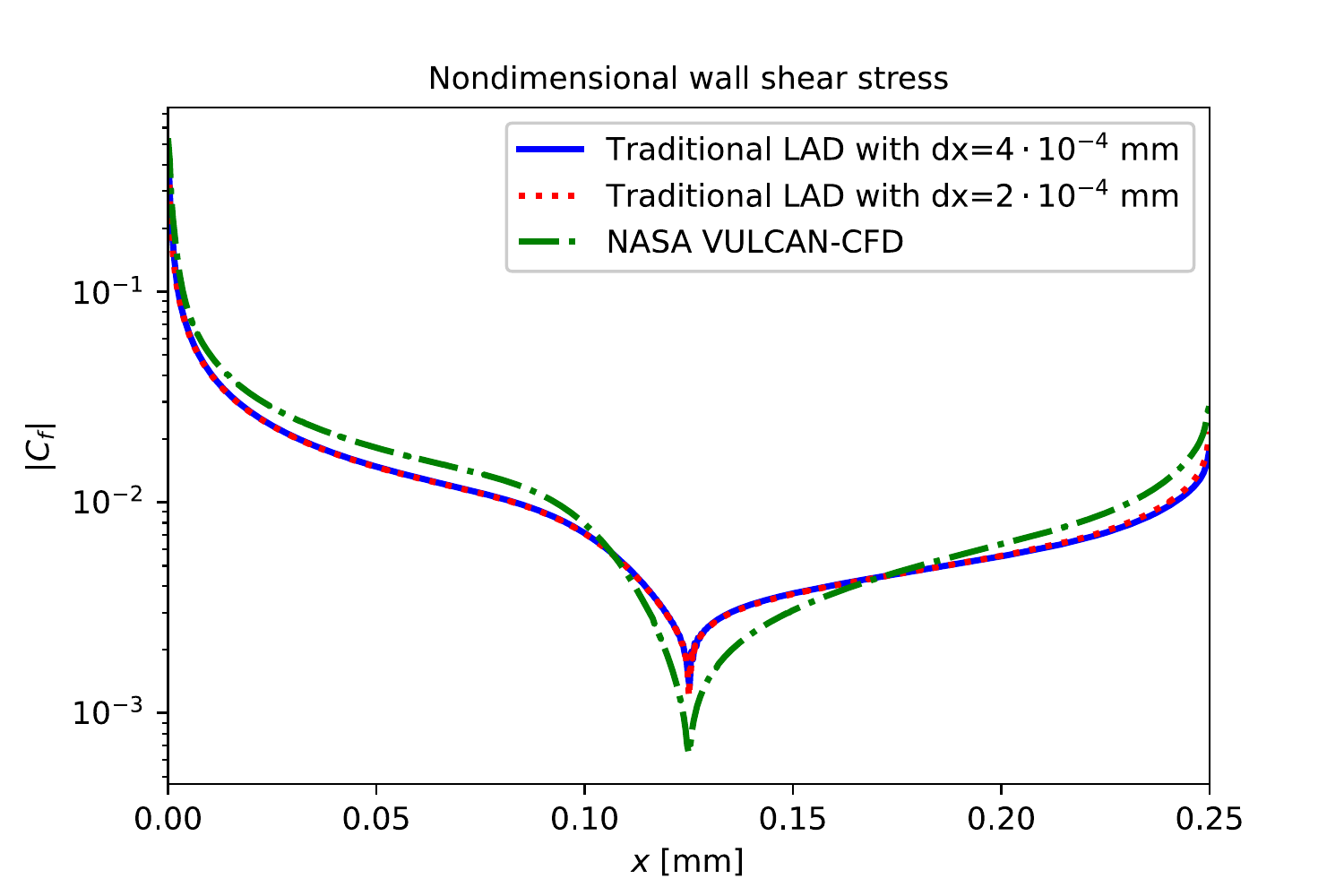}
  \caption{}
  \label{fig:C_f_12deg}
  \end{subfigure}
         \begin{subfigure}{0.49 \textwidth}
\includegraphics[width=0.95\textwidth]{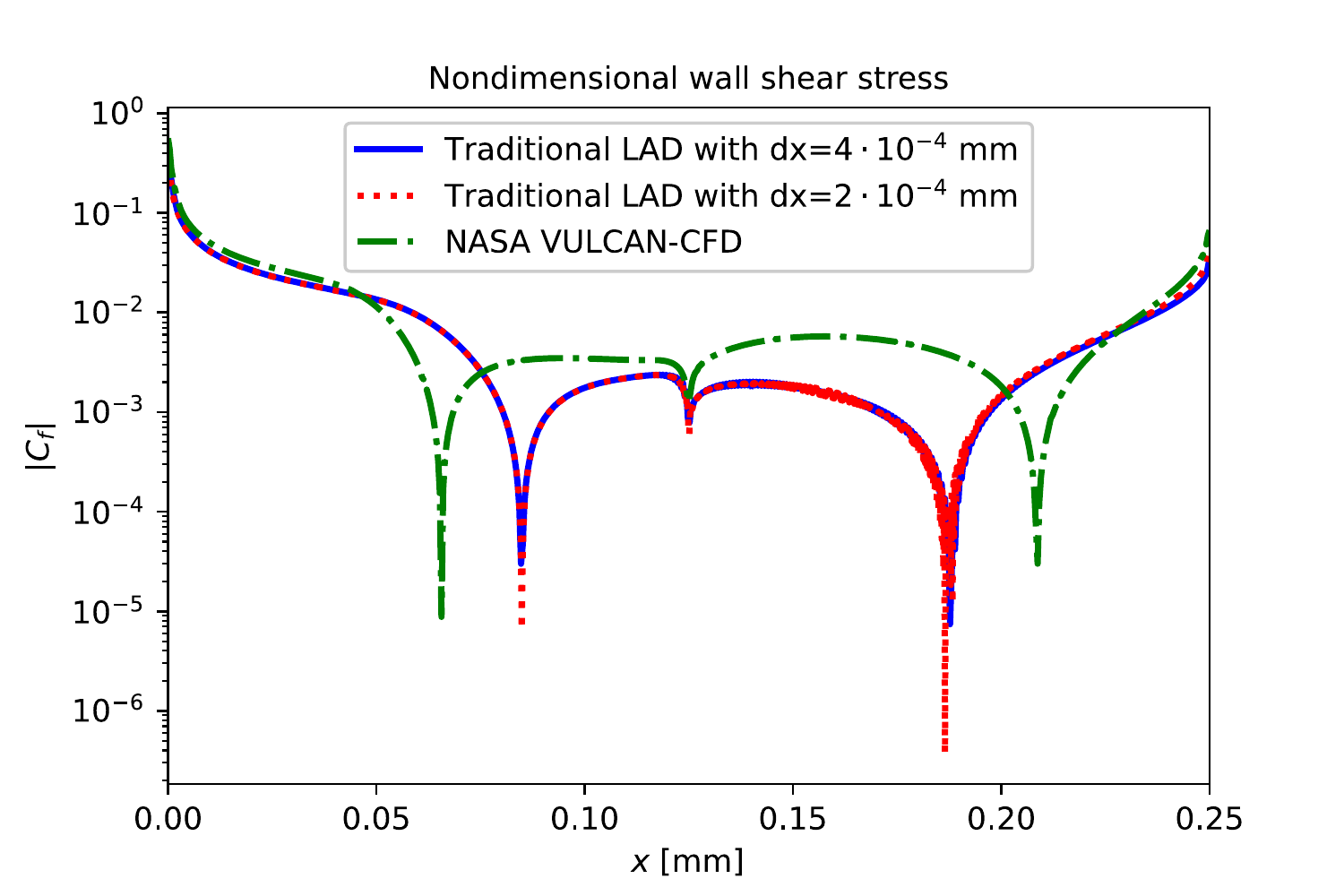}
  \caption{}
  \label{fig:C_f_24deg}
  \end{subfigure}
    \caption{Comparison of key wall metrics between traditional LAD and \vulkan: non-dimensional wall heat flux for (a) $\alpha=12^\circ$ (a) and (b) $24^\circ$; non-dimensional wall pressure for (c) $\alpha=12^\circ$ and (d) $24^\circ$; and non-dimensional wall shear stress for (e) $\alpha=12^\circ$ and (f) $24^\circ$.}
    \label{fig:soa_lad_vulcan_comp}
\end{figure}

We consider flow turning (deflection) angles $\alpha=12^\circ$, involving no flow separation or shock merging, and $\alpha=24^\circ$, for which flow separation and recirculation are expected, along with shock merging. For each case, we compare the predictions of traditional LAD using second-order CD with those of NASA's \vulkan code \cite{NASA2021} in terms of three important wall metrics; namely, non-dimensional heat flux, pressure, and shear stress (Figure \ref{fig:soa_lad_vulcan_comp}).%

For the current problem, we choose $l$ to be $1/2$, $0$, $0$ and $1/2$, in Eq.~\eqref{eqn:AMD_Fiorina}, Eq.~\eqref{eqn:ASV}, Eq.~\eqref{eqn:ABV_Mani} and Eq.~\eqref{eqn:AHC_Cook}respectively, and set the user-defined constants ${C}_{\chi}^{1/2}$, ${C}_{\mu}^{0}$, ${C}_{\beta}^{0}$ and ${C}_{k}^{1/2}$ equal to:
\begin{align}
   {C}_{\chi}^{1/2} &:= \hat{C}_{\chi}\frac{c_P}{a_0\Delta}\frac{\|\vec{u}\|_{\text{max}}}{\|\nabla s\|_{\text{max}}}, \\
  {C}_{\mu}^{0} &:= \hat{C}_\mu \frac{\rho_{\text{max}}|(\epsilon:\epsilon)^\frac{1}{2}|_{\text{max}}}{\big[\rho|\nabla\cdot\vec{u}|H(-\nabla\cdot\vec{u})\big]_{\text{max}}},\\
  {C}_{\beta}^{0} &:= \frac{\hat{C}_\beta}{\Delta}  \frac{\rho_{\text{max}}\|\vec{u}\|_{\text{max}}}{\big[\rho|\nabla\cdot\vec{u}|H(-\nabla\cdot\vec{u})\big]_{\text{max}}}, \\
  {C}_{k}^{1/2} &:= \frac{\hat{C}_k}{a_0\Delta} \frac{R\rho_{\text{max}}\|\vec{u}\|_{\text{max}}}{(\gamma-1)\big[(\rho/T) \|\nabla e\|\big]_{\text{max}}}.
\end{align}
Here we set $\hat{C}_{\chi}:=0$, $\hat{C}_\mu:=0.02$, $\hat{C}_\beta:=0.04$ and $\hat{C}_{k}:=0.02$ \footnote{These values are based on a limited parameter sweep and aim to keep the maximum artificial dissipation as low as possible, while still maintaining a stable simulation. A more extensive parameter sweep may yield more optimal values, but would lead to additional computational overhead.} Moreover, we consider computational meshes with grid cell dimensions of $4\times 10^{-7}$ m and $2\times 10^{-7}$ m in the horizontal and vertical directions. The grids are slightly nonuniform near the wall because of the corner geometry, causing the exact grid cell dimensions in that region to vary somewhat from the specified values.

We performed all simulations of the compression corner in \openfoam. While the derivation of the artificial PDE terms took a few months, thanks to the use of CD for the spatial discretization, the numerical source code implementing the modified PDE approach was straightforward to write. Moreover, as part of the CompMods final deliverables, we illustrated how a ``multi-physics compiler'' could automate the generation of this source code, demonstrating the power of IAT to produce directly computable models.

\begin{figure}
\centering
\begin{subfigure}{0.46 \linewidth}
\includegraphics[width=0.95\textwidth]{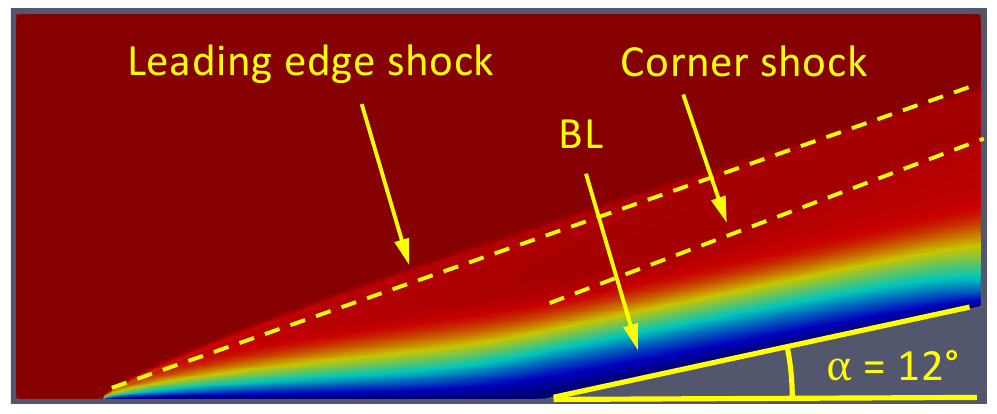}
  \caption{}
\end{subfigure}
\begin{subfigure}{0.53 \linewidth}
\includegraphics[width=0.95\textwidth]{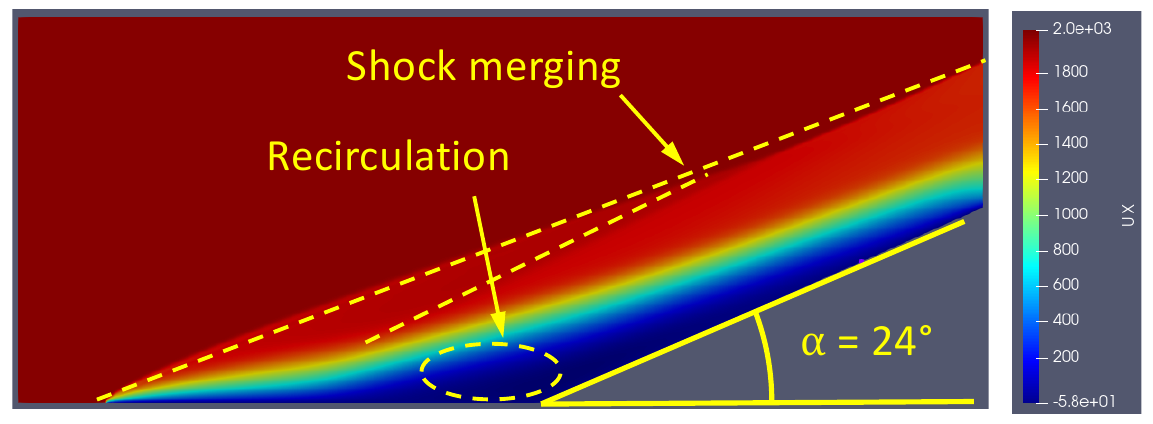}
  \caption{}
  \end{subfigure}
    \caption{Visualization of the horizontal velocity field for turning angles $\alpha=12^\circ$ (left) and $\alpha=24^\circ$ (right) near steady-state, obtained with traditional LAD. Various flow features are indicated, illustrating the change in flow regime when going to higher corner angles.}
      \label{fig:corner_visualization_vel}
\end{figure}

\begin{figure}
\centering
\begin{subfigure}{0.48 \linewidth}
\includegraphics[width=0.95\textwidth]{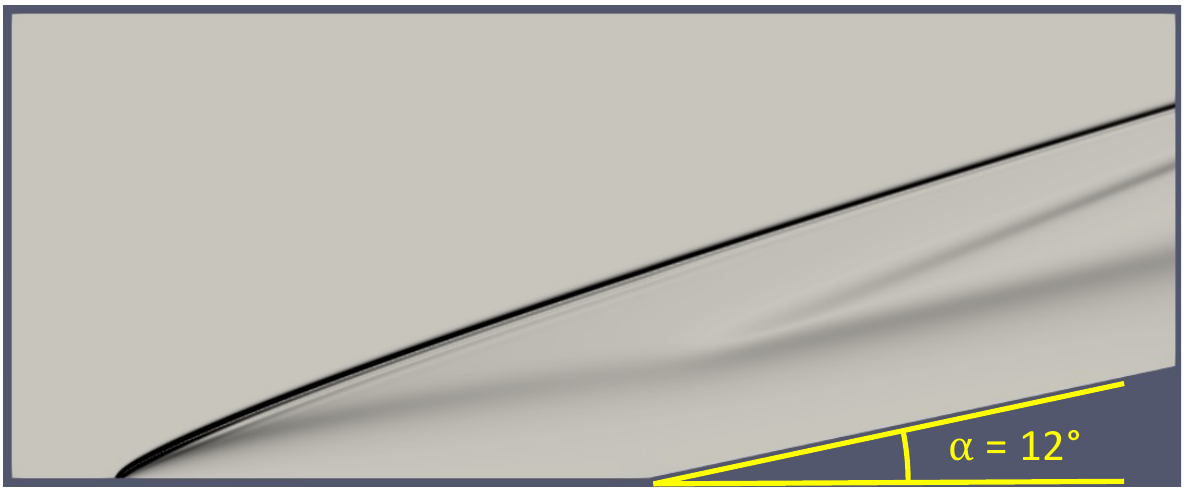}
  \caption{}
\end{subfigure}
\begin{subfigure}{0.49 \linewidth}
\includegraphics[width=0.95\textwidth]{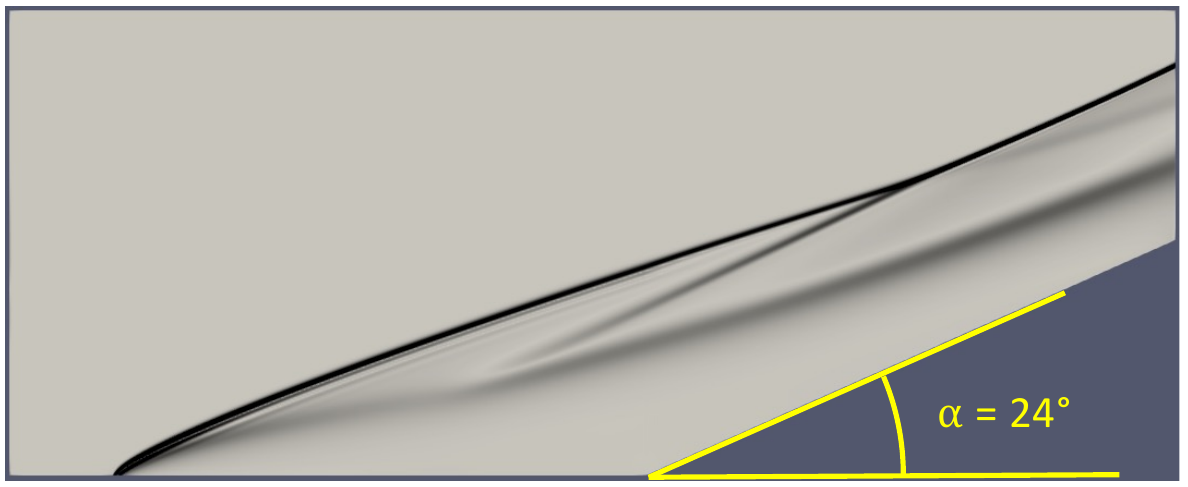}
  \caption{}
  \end{subfigure}
    \caption{Visualization of the magnitude of the density gradient for turning angles $\alpha=12^\circ$ (left) and $\alpha=24^\circ$ (right) near steady-state, obtained with traditional LAD.}
      \label{fig:corner_visualization_gradrho}
\end{figure}

Figures \ref{fig:corner_visualization_vel} and \ref{fig:corner_visualization_gradrho} visualize the horizontal velocity field and magnitude of the density gradient, respectively, near steady-state for the 12$^\circ$ (left) and 24$^\circ$ (right) turning angles. We can clearly see the difference in flow regime, with the flow staying attached to the wall and non-intersecting leading edge and corner shocks for the smaller 12$^\circ$ angle, and flow separation and recirculation with merging of the leading edge and corner shocks for the larger 24$^\circ$ angle. Steady-state is reached much sooner in the former case (shown at $t=8\times 10^{-7}$ s) than in the latter case (shown at $t=1.6\times 10^{-6}$ s).

For both investigated turning angles of $\alpha=12^\circ$ and $\alpha=24^\circ$, Figure \ref{fig:soa_lad_vulcan_comp} shows that LAD is able to capture the overall shape and magnitude of the spatial profiles for the three wall metrics.
Furthermore, as Table \ref{tab:rel_err} shows, the wall-integrated heat flux, pressure and shear stress predicted by the LAD model using a grid cell size of $4\times 10^{-7}$ m stays below 12\% and 10\% in relative error with respect to \vulkan for $\alpha=12^\circ$ and $\alpha=24^\circ$, respectively.
\begin{table}
  \centering
  \caption{Relative error in the wall-integrated heat flux, pressure and shear stress computed from the spatial profiles of the wall metrics in Figure \ref{fig:soa_lad_vulcan_comp} between traditional LAD and \vulkan.}
  \label{tab:rel_err}
  \begin{tabular}{cccccc}
    \toprule%
    \multicolumn{1}{l}{Turning angle} & Total heat & Pressure force & Shear force & Pressure force & Shear force\\%
    \multicolumn{1}{l}{\ \ \ \ \ \ \ ($\alpha$)} & flux & along $x$ (drag) & along $x$ (drag) & along $y$ (lift) & along $y$ (lift)\\%
    \midrule%
    $12^\circ$  & 11.31\% & 2.95\% & 8.80\% & 11.30\% & 0.06\% \\%
    $24^\circ$  & 9.70\% & 5.04\% & 0.16\% & 5.91\% & 0.14\% \\%
    \bottomrule
  \end{tabular}
\end{table}
However, at the larger angle of $24^\circ$, we note the presence of nonphysical oscillations in the wall pressure predicted by LAD, as well as the substantial deviation of its predicted locations of separation and reattachment from their \vulkan counterparts, as can be seen from the wall shear stress profiles in Figure \ref{fig:soa_lad_vulcan_comp}(f). As Figure \ref{fig:soa_lad_vulcan_comp} demonstrates, increasing the mesh resolution did not improve the agreement of the LAD-predicted wall shear stress with its \vulkan counterpart, and further increased the amplitude of the oscillations in the wall pressure. While further tuning of the artificial parameters $\hat{C}_\chi$, $\hat{C}_\mu$, $\hat{C}_\beta$ and $\hat{C}_k$ could improve the match between the LAD and \vulkan results, this reveals the inherent problems caused by having tunable parameters in the model. A tuning-free approach that rectifies these shortcomings is presented next.

\subsection{Limiter-inspired LAD: LAD inspired by TVD-endowing flux limiters}
\label{subsec:TVD_LAD}

As described in Section \ref{subsubsec:results_1D_trad_LAD}, combining LAD in its traditional formulation with second-order CD may result in excessive undershoots/overshoots as well as numerical diffusion, and requires problem-dependent parameter tuning. This shortcoming motivates exploration of a novel LAD approach inspired by flux limiters endowing certain high-resolution schemes like MUSCL with the total variation diminishing (TVD) property that aims to achieve oscillation-free solutions to shocked compressible flow problems. Such an approach is an example of IAT via PDE modifications derived by ``reverse engineering'' numerical methods. Below we develop our approach for the case of a 1D uniform grid, and discuss its extension to multi-dimensional domains in Section \ref{subsec:multiD_TVD_insp_LAD}.

\begin{figure}
\centering
\begin{subfigure}{0.47 \linewidth}
\includegraphics[width=0.9\textwidth]{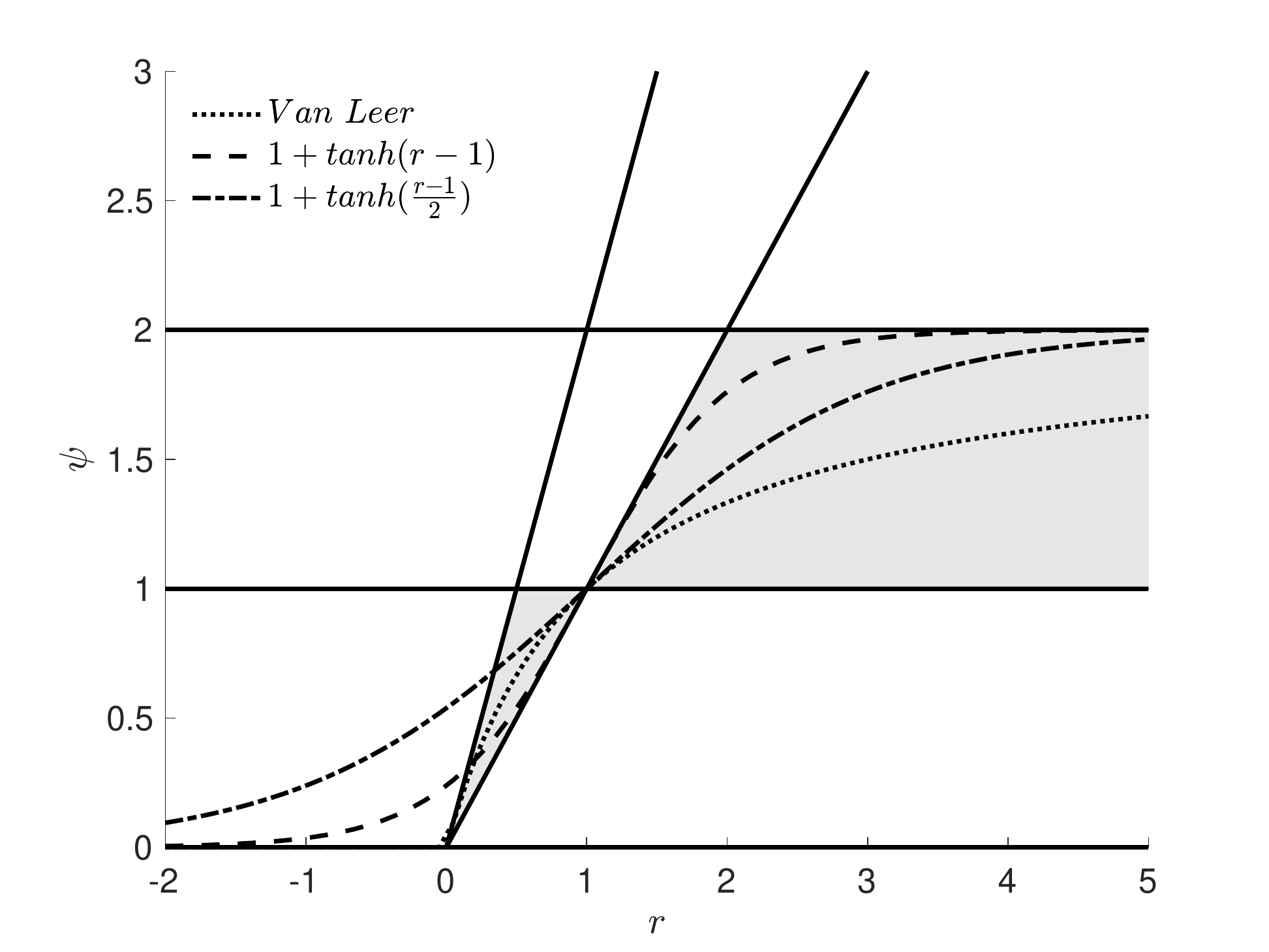}
  \caption{}
\end{subfigure}
\begin{subfigure}{0.50 \linewidth}
\includegraphics[width=0.95\textwidth]{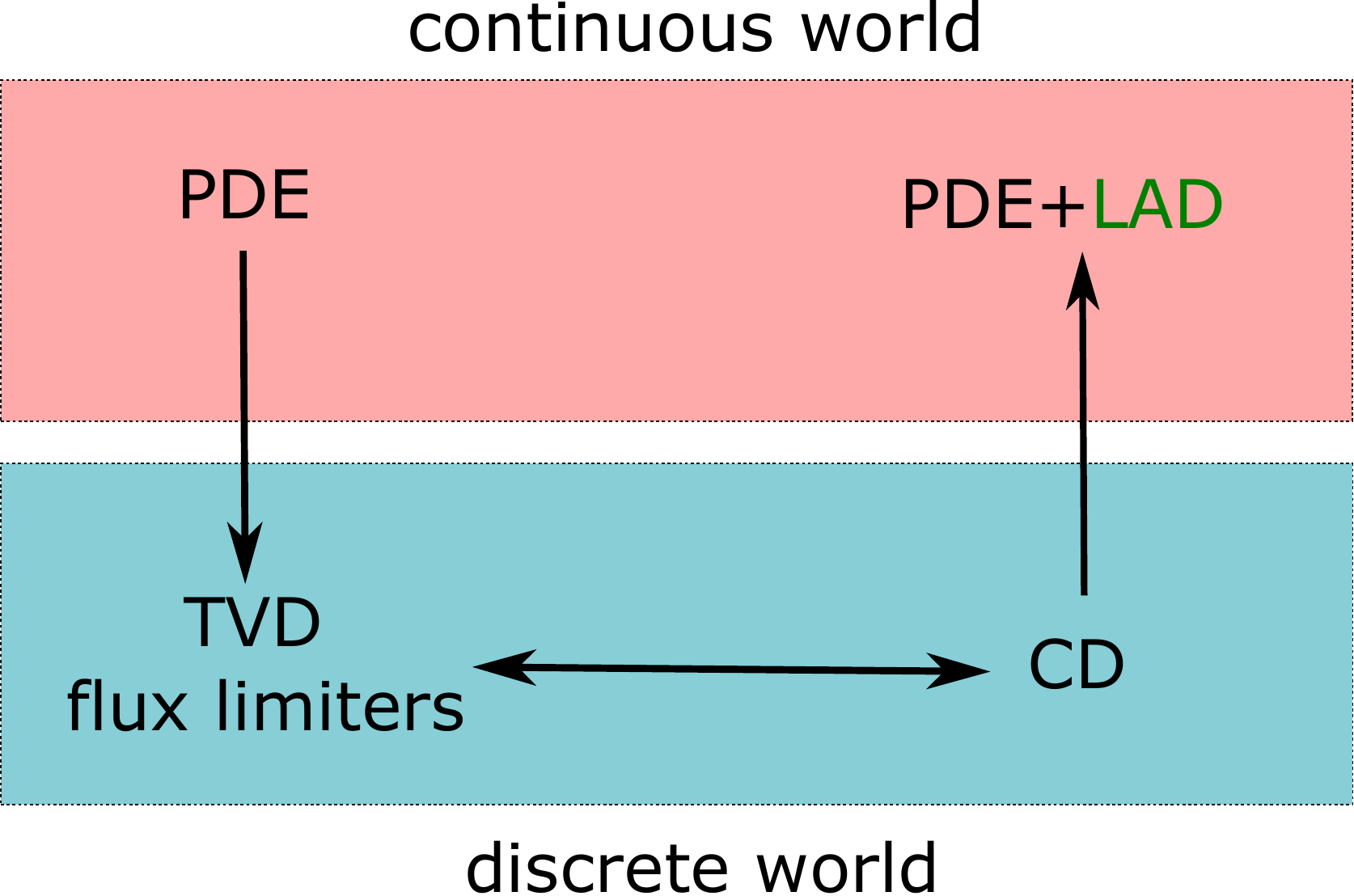}
  \caption{}
  \end{subfigure}
    \caption{(a) The $r$-$\psi$ parameter space, where the TVD region is shaded, in addition to plots of the curves corresponding to the van Leer flux limiter and the limiter-inspired LAD for two different hyperparameter values. (b) A schematic of the approach for deriving the limiter-inspired LAD model.}
      \label{fig:TVD+schematic}
\end{figure}

Suppose that we are advecting a scalar $\phi$ with positive velocity ($u > 0$), along one dimension. This is described by the PDE:
\begin{equation}
    \frac{\partial \phi}{\partial t}+\frac{\partial f}{\partial x} = 0,
    \label{eqn:PDE_pure}
\end{equation}
where $f\equiv \phi u$ is the advective flux. Upon discretization of the domain, the flux on the right face of the $i^\text{th}$ cell, denoted by $f_{i+\nicefrac{1}{2}}$, can be expressed as:
\begin{equation}
    f_{i+\nicefrac{1}{2}}=f_i+\frac{1}{2}\psi(r)(f_{i+1}-f_{i}),
    \label{eqn:TVD_definition}
\end{equation}
where $r$ is the ratio of the upwind-side gradient to the downwind-side gradient of the flux, defined as:
\begin{equation}
    r=\frac{f_i-f_{i-1}}{f_{i+1}-f_i},
    \label{eqn:r_definition}
\end{equation}
and $\psi(r)$ is the {\it flux limiter} function that determines the type of scheme. For example, $\psi(r) := (r+3)/4$ corresponds to the QUICK scheme \cite{Leonard1979} and $\psi(r) :=(|r|+r)/(1+r)$ corresponds to the van Leer scheme \cite{vanLeer1974}. Other famous second-order flux limiters include minmod \cite{Roe1986}, Superbee \cite{Roe1986}, and van Albada \cite{VanAlbada}. Any valid second-order accurate flux limiter needs to satisfy a number of conditions, namely:
\begin{enumerate}
    \item $\psi(r=1)=1$, i.e., the scheme reverts to second-order CD, in smooth parts of the solution \cite{Sweby1984}.
    \item For $r>0$, $\psi(r)$ should fall in the shaded region in Figure \ref{fig:TVD+schematic}(a) to ensure the TVD property is satisfied \cite{Sweby1984}.
    \item For $r<0$, $\psi(r)=0$, i.e., the scheme applies first-order upwind differencing (UD) at local extrema.
\end{enumerate}
The key idea is to find the PDE modification for which, upon applying a second-order CD discretization on the resulting modified PDE, the discrete equations\footnote{We are only referring to spatial discretization here.} would be equivalent to those obtained using a TVD-endowing flux limiter on the original PDE. With this insight, we can identify the main ingredients required to achieve robust, oscillation-free solutions, based on which we propose a PDE modification in the form of a simple artificial diffusivity localization which is second-order, quasi-TVD, and tuning-free. A schematic of this approach is shown in Figure \ref{fig:TVD+schematic}(b).

From here on, we consider the grid to be uniform for ease of notation, but the extension to nonuniform grids is straightforward. For example, let us consider a first-order UD scheme (i.e., $f_{i+\nicefrac{1}{2}} :=f_i$). The second-order CD operator for the first derivative is defined as:
\begin{equation}
    \frac{\partial f}{\partial x}\Big\vert_{i+\nicefrac{1}{2}}^{\rm CD} := \frac{f_{i+1}-f_{i}}{\Delta x},
    \label{eqn:first_derivative}
\end{equation}
and for the second derivative is obtained as:
\begin{equation}
    \frac{\partial^2 f}{\partial x^2}\Big\vert_{ i}^{\rm CD} := \frac{\partial}{\partial x} \Big\vert_{ i}^{\rm CD} \frac{\partial f}{\partial x}\Big\vert_{i}^{\rm CD} = \frac{\frac{\partial f}{\partial x}\vert_{i+\nicefrac{1}{2}}^{\rm CD}-\frac{\partial f}{\partial x}\vert_{ i-1/2}^{\rm CD}}{\Delta x} = \frac{\frac{f_{i+1}-f_{i}}{\Delta x} - \frac{f_{i}-f_{i-1}}{\Delta x}}{\Delta x} = \frac{f_{i+1}-2f_{i}+f_{i-1}}{\Delta x^2},
    \label{eqn:second_derivative}
\end{equation}
along with linear (i.e., accurate up to the second-order) interpolation, obtained as:
\begin{equation}
    {f}\Big\vert_{i+\nicefrac{1}{2}}^{\rm CD}=\frac{f_{i+1}+f_{i}}{2},
    \label{eqn:interpolation}
\end{equation}
where $\Delta x := (x_{i+1} - x_i) = (x_i - x_{i-1})$. The first-order UD operator can then be expressed as:
\begin{equation}
    f_{i+\nicefrac{1}{2}}\Big\vert_{i+\nicefrac{1}{2}}^{\rm UD}={f}\Big\vert_{i+\nicefrac{1}{2}}^{\rm CD}-\frac{\Delta x}{2}\frac{\partial f}{\partial x}\Big\vert_{i+\nicefrac{1}{2}}^{\rm CD}=(f-\frac{\Delta x}{2}\frac{\partial f}{\partial x})\Big\vert_{i+\nicefrac{1}{2}}^{\rm CD}.
    \label{eqn:upwind_flux}
\end{equation}
Thus, using first-order UD to compute the advective face fluxes for Eq.~\eqref{eqn:PDE_pure} is equivalent to using second-order CD to discretize the modified PDE in which the right-hand side zero is replaced with an artificial flux divergence:
\begin{equation}
    \frac{\partial \phi}{\partial t}+\frac{\partial f}{\partial x}=\frac{\partial}{\partial x}(D^\ast\frac{\partial f}{\partial x}),
    \label{eqn:TVD_PDE_Model}
\end{equation}
in which the artificial diffusivity is $D^\ast = \Delta x/2$ to match Eq.~\eqref{eqn:upwind_flux} upon discretization with second-order CD. This is a well-known result stating that first-order UD is equivalent to applying second-order CD in addition to mesh-dependent artificial diffusion. Note that diffusion is operating on the flux, not the primitive variable $\phi$, hence it has units that are different from those of regular diffusivities. Analogously, instead of computing advective face fluxes using Eq.~\eqref{eqn:TVD_definition}, the modified PDE of Eq.~\eqref{eqn:TVD_PDE_Model} can be discretized using second-order CD, where the artificial diffusivity $D^\ast$ is given by:
\begin{equation}
    D^\ast = \frac{\Delta x}{2}\text{sgn}(u) \big( 1-\psi(r) \big),
    \label{eqn:D_TVD}
\end{equation}
with $\text{sgn}(u)$ incorporating the direction of the flow.

To have a well-defined modification to the PDE, $\psi(r)$ must be computed locally over the mesh faces. While $\psi(r)$ is a known analytical function, its argument $r$ should be found as a composition of continuous operations. To compute the non-local fluxes involved in the definition of $r$ in Eq.~\eqref{eqn:r_definition} in terms of second-order CD operations at $x_{i+\nicefrac{1}{2}}$, we use Eqs.~\eqref{eqn:first_derivative}--\eqref{eqn:interpolation} in addition to a filtering operation:
\begin{equation}
    \overline{f}_i=\frac{1}{4}f_{i-1}+\frac{1}{2}f_{i}+\frac{1}{4}f_{i+1},
\end{equation}
to obtain:
\begin{equation}
    f_i ~= \left(f-\frac{\Delta x}{2}\frac{\partial f}{\partial x}\right)\Big\vert_{i+\nicefrac{1}{2}}^{\rm CD},
    \label{eqn:f_i_central}
\end{equation}
\begin{equation}
    f_{i+1}=\left(f+\frac{\Delta x}{2}\frac{\partial f}{\partial x}\right)\Big\vert_{i+\nicefrac{1}{2}}^{\rm CD},
    \label{eqn:f_i+1_central}
\end{equation}
\begin{equation}
    f_{i-1}=\left(f+\frac{\Delta x}{2}\frac{\partial f}{\partial x}-2 \Delta x \frac{\partial \overline{f}}{\partial x} + (\Delta x)^2\frac{\partial^2 f}{\partial x^2}\right)\Big\vert_{i+\nicefrac{1}{2}}^{\rm CD}.
    \label{eqn:f_i-1_central}
\end{equation}
Substituting Eqs.~\eqref{eqn:f_i_central}--\eqref{eqn:f_i-1_central} into Eq.~\eqref{eqn:r_definition} yields:
\begin{equation}
    r = \left(\frac{2\frac{\partial \overline{f}}{\partial x}-\frac{\partial f}{\partial x} - \Delta x ~\text{sgn}(u)\frac{\partial^2 f}{\partial x^2}}{\frac{\partial f}{\partial x}}\right)\Big\vert_{i+\nicefrac{1}{2}}^{\rm CD}.
\label{eqn:PDE_r}
\end{equation}
Thus, for example, for the van Leer limiter $\psi(r) := (|r|+r)/(1+r)$, we find that the artificial diffusivity $D^\ast$ is given by:
\begin{equation}
    D^\ast = \frac{\Delta x}{2}\text{sgn}(u)\frac{\frac{1}{2}\frac{\partial f}{\partial x}-\frac{\partial f/\partial x}{|\partial f/\partial x|}|\frac{\partial \overline{f}}{\partial x}-\frac{\Delta x}{2}\text{sgn}(u)\frac{\partial^2 f}{\partial x^2}-\frac{1}{2}\frac{\partial f}{\partial x}|}{\frac{\partial \overline{f}}{\partial x}-\frac{\Delta x}{2}\text{sgn}(u)\frac{\partial^2 f}{\partial x^2}}.
    \label{eqn:VanLeer_D}
\end{equation}
Similarly, one can find the diffusivity expressions for other flux limiter functions such as minmod and Superbee. While being equivalent to applying a TVD-endowing flux limiter on the original PDE, this “limiter-based” LAD has the benefit of being able to use central spatial operators to achieve TVD properties, rendering this strategy accessible to common multi-physics software frameworks like \openfoam.

Although this limiter-based approach seems reasonable, its formulation can get very complicated depending on the flux limiter being considered. Recognizing that respecting the important limits of (a) $\psi(r=1)=1$, (b) $\psi(r\rightarrow -\infty)=0$, and (c) $1\le\psi(r\rightarrow +\infty)\le2$, maintains the robustness of these flux limiters despite only approximately satisfying the TVD property, we go one step further and devise a “limiter-inspired” LAD scheme that avoids excessive diffusion while using a simpler formulation. One option is to use:
\begin{equation}
    \psi(r) := 1+\text{tanh}\left(\frac{r-1}{\kappa}\right),
    \label{eqn:tanh_formula}
\end{equation}
where $\kappa$ is a hyperparameter that sets the width of the hyperbolic tangent transition. From Eqs.~\eqref{eqn:D_TVD} and \eqref{eqn:PDE_r}, we can see that this corresponds to setting $D^\ast$ equal to:
\begin{equation}
    D^\ast = \frac{\Delta x}{2}\text{sgn}(u)\text{tanh}\left[2\frac{\frac{\partial f}{\partial x}-\frac{\partial \overline{f}}{\partial x}+\frac{\Delta}{2}\text{sgn}(u)\frac{\partial^2 f}{\partial x^2}}{\kappa\frac{\partial {f}}{\partial x}}\right].
    \label{eqn:TVD-inspired_D}
\end{equation}
In order to avoid a negative artificial viscosity or heat conductivity and to reduce oscillations, we clip and filter the computed diffusivity values, i.e., we use
\begin{equation}
    D^\ast = G\left({\max\left\{\frac{\Delta x}{2}\text{sgn}(u)\text{tanh}\left[2\frac{\frac{\partial f}{\partial x}-\frac{\partial \overline{f}}{\partial x}+\frac{\Delta x}{2}\text{sgn}(u)\frac{\partial^2 f}{\partial x^2}}{\kappa\frac{\partial {f}}{\partial x}}\right],0\right\}}\right),
    \label{eqn:TVD-inspired_D_clipped_filtered}
\end{equation}
where $G$ is a Gaussian filter. Acknowledging that there is no physical counterpart to artificial mass diffusion (AMD), we use Eq.~\eqref{eqn:TVD-inspired_D} for AMD to maintain sharp discontinuities (i.e., we allow for negative artificial mass diffusivity). Figure \ref{fig:TVD+schematic}(a) depicts the $\psi(r)$ curve for $\kappa=1,2$, from which it is clear that both of these curves lie in the TVD region for most values of $r$. While $\kappa=1$ guarantees the TVD property for a slightly larger set of $r$ values, $\kappa=2$ is preferred because it is less diffusive and sufficiently dampens oscillations due to CD. In other words, we sacrifice exact satisfaction of the TVD property in order to reduce artificial diffusion of physical oscillations while maintaining robustness of the solutions (i.e., avoiding numerical oscillations). All results in this work obtained using Eqs.~\eqref{eqn:TVD-inspired_D}--\eqref{eqn:TVD-inspired_D_clipped_filtered} therefore employ $\kappa=2$. No problem-dependent tuning is required for this hyperparameter, making the limiter-inspired approach completely free of problem-specific tunable parameters. In principle, once a PDE modification is found, it allows for the use of spatial schemes of various orders of accuracy, and for the use of different numerical schemes for different terms.

\subsubsection{Verification of IAT requirements}
\label{subsubsec:verification_IAT_LI_LAD}
Since the limiter-inspired LAD approach augments the compressible flow equations similar to the traditional LAD schemes (Eqs.~\eqref{eqn:mass_PDE}-\eqref{eqn:energy_PDE}), verifying the IAT requirements is quite straightforward. In particular, the first requirement is satisfied because the governing equations are the same, satisfying conservation principles and thermodynamic consistency. According to Eq.~\eqref{eqn:D_TVD} and the $O(1)$ values for $\psi(r)$ (see Figure \ref{fig:TVD+schematic}), for the diffusivity of the limiter-inspired LAD model, we have $D^\ast\sim\Delta$. Recall that $D^\ast$ operates on fluxes. Taking into account the velocity-dependent scaling of the fluxes, it is clear that the second, third, and fourth requirements are satisfied based on the same reasoning as for traditional LAD schemes. The fifth requirement is satisfied because, in the smooth regions away from discontinuities, $\psi$ is designed to go to $1$, resulting in no artificial diffusivity. As such, the original equations are recovered away from the discontinuities and the Rankine-Hugoniot jump conditions reveal that the speed of the discontinuities is unaltered by the localized diffusion. Finally, the sixth requirement is satisfied because as the mesh is made finer, the artificial diffusivity shrinks and the solution to the augmented equations converges to the entropy solution satisfying the entropy condition.

\subsubsection{1D canonical problems}
\label{sec:results_tvd_inspired}

\begin{figure}
\centering
\begin{subfigure}{0.49 \textwidth}
\includegraphics[width=0.95\textwidth]{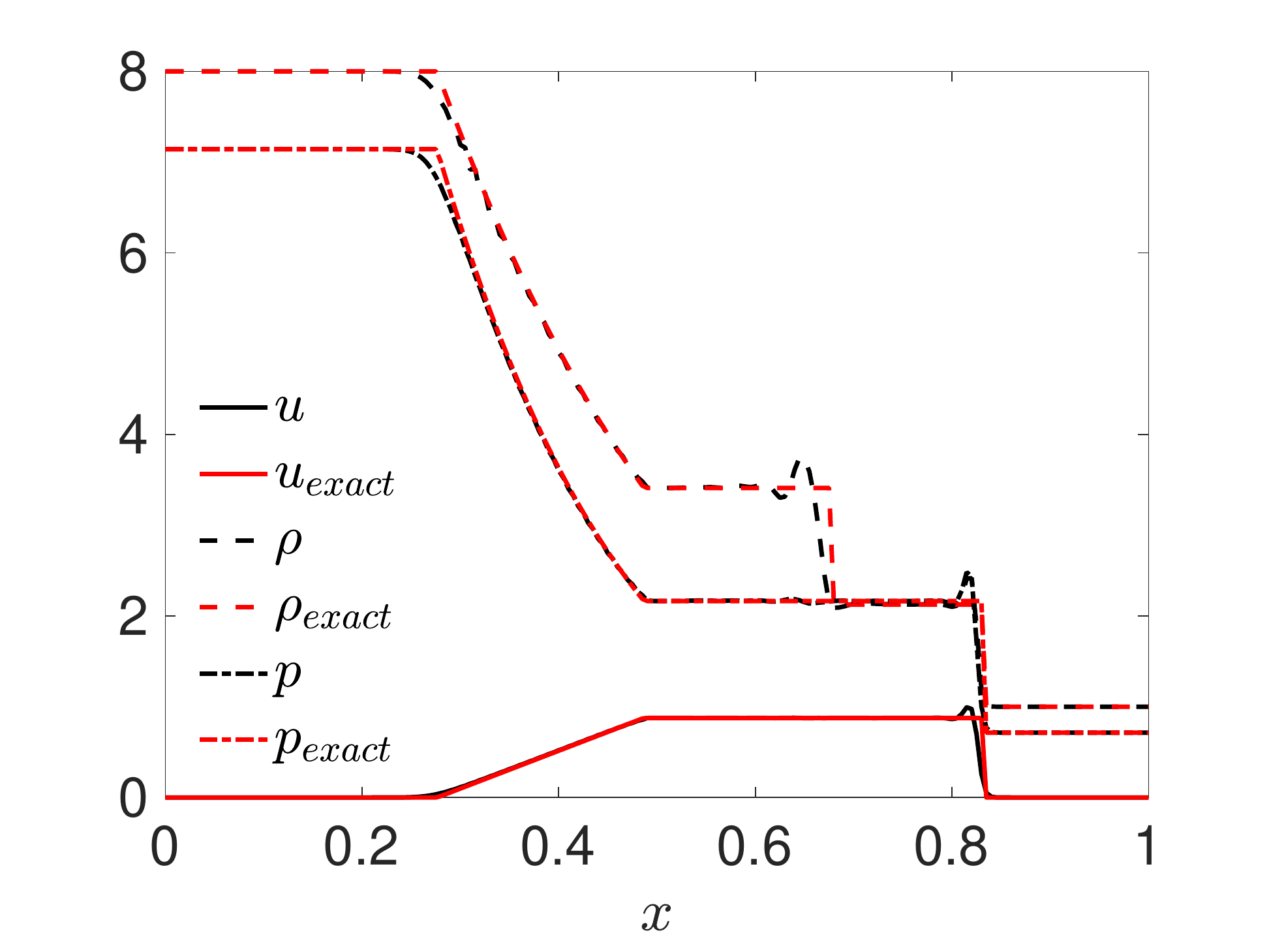}
  \caption{}
\end{subfigure}
\begin{subfigure}{0.49 \textwidth}
\includegraphics[width=0.95\textwidth]{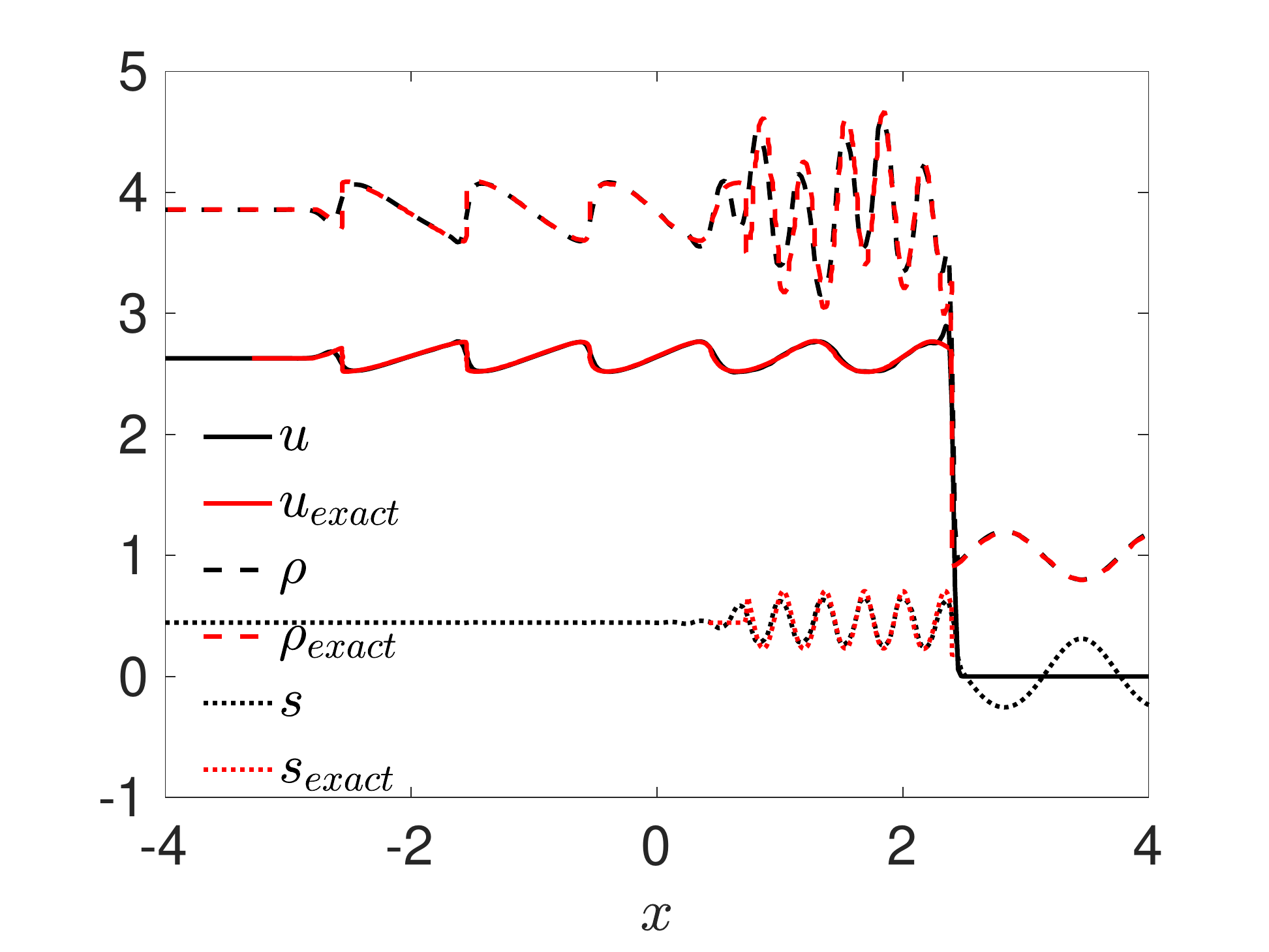}
  \caption{}
  \end{subfigure}
         \begin{subfigure}{0.49 \textwidth}
\includegraphics[width=0.95\textwidth]{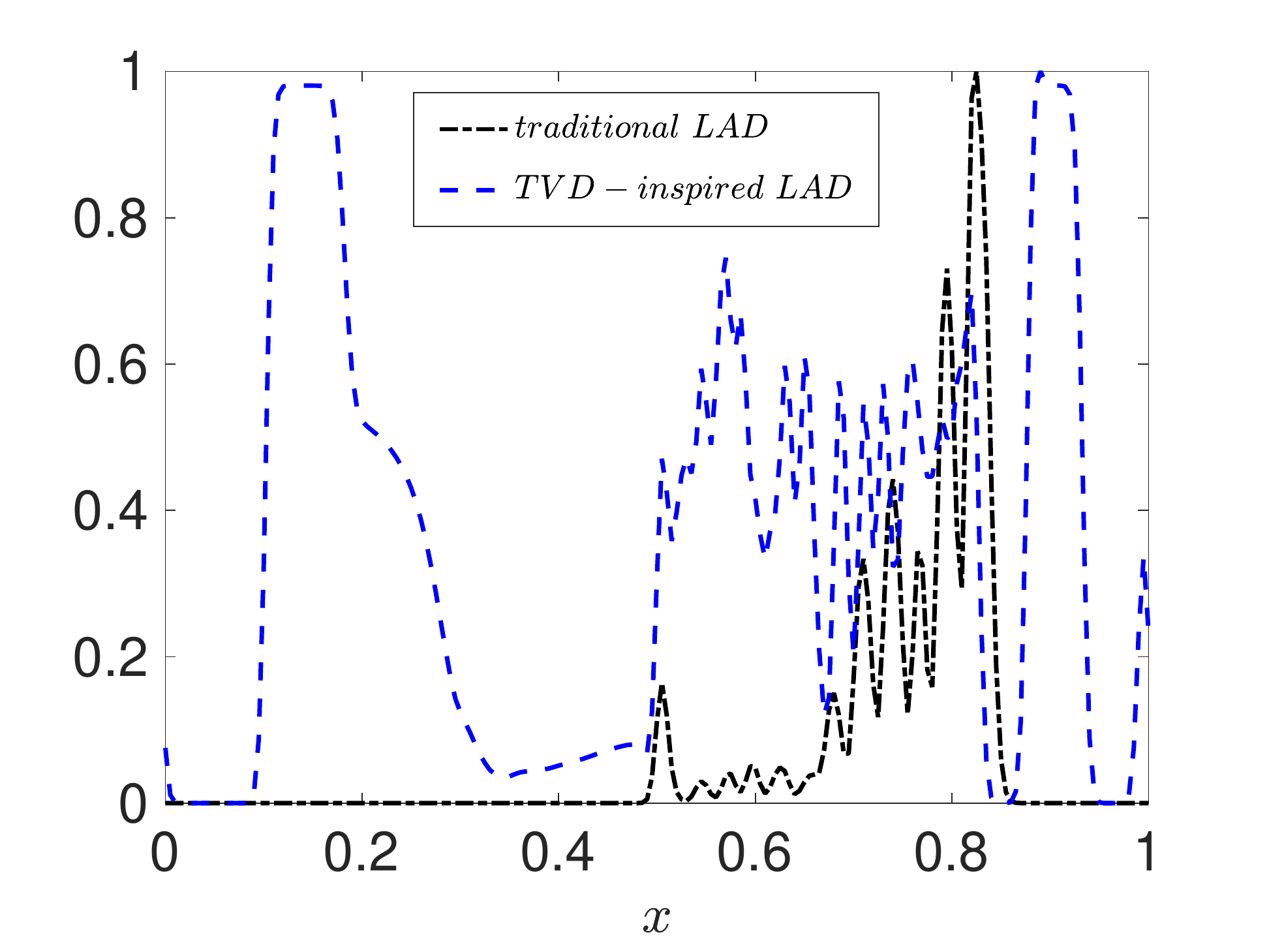}
  \caption{}
  \end{subfigure}
         \begin{subfigure}{0.49 \textwidth}
\includegraphics[width=0.95\textwidth]{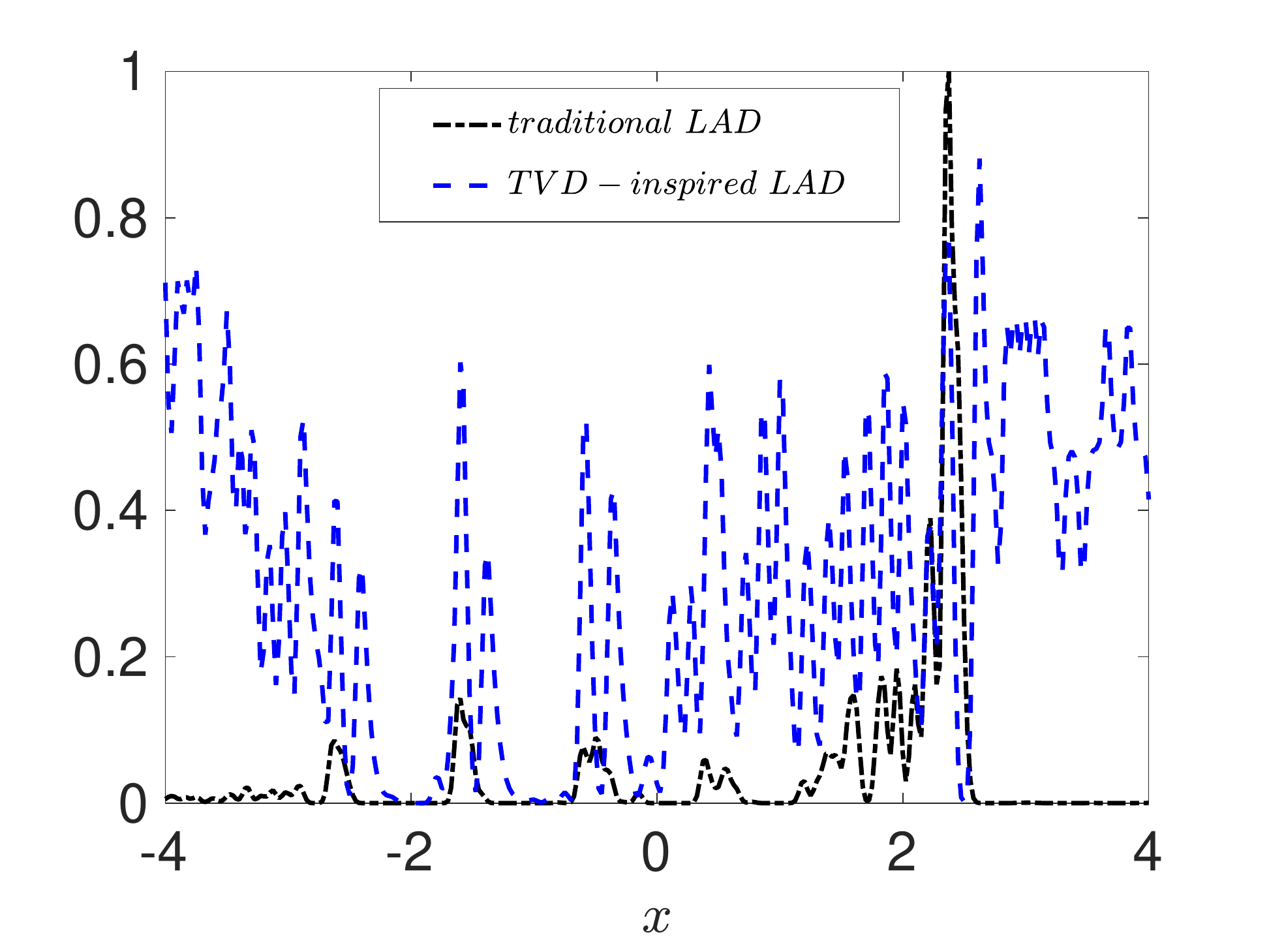}
  \caption{}
  \end{subfigure}
           \begin{subfigure}{0.49 \textwidth}
\includegraphics[width=0.95\textwidth]{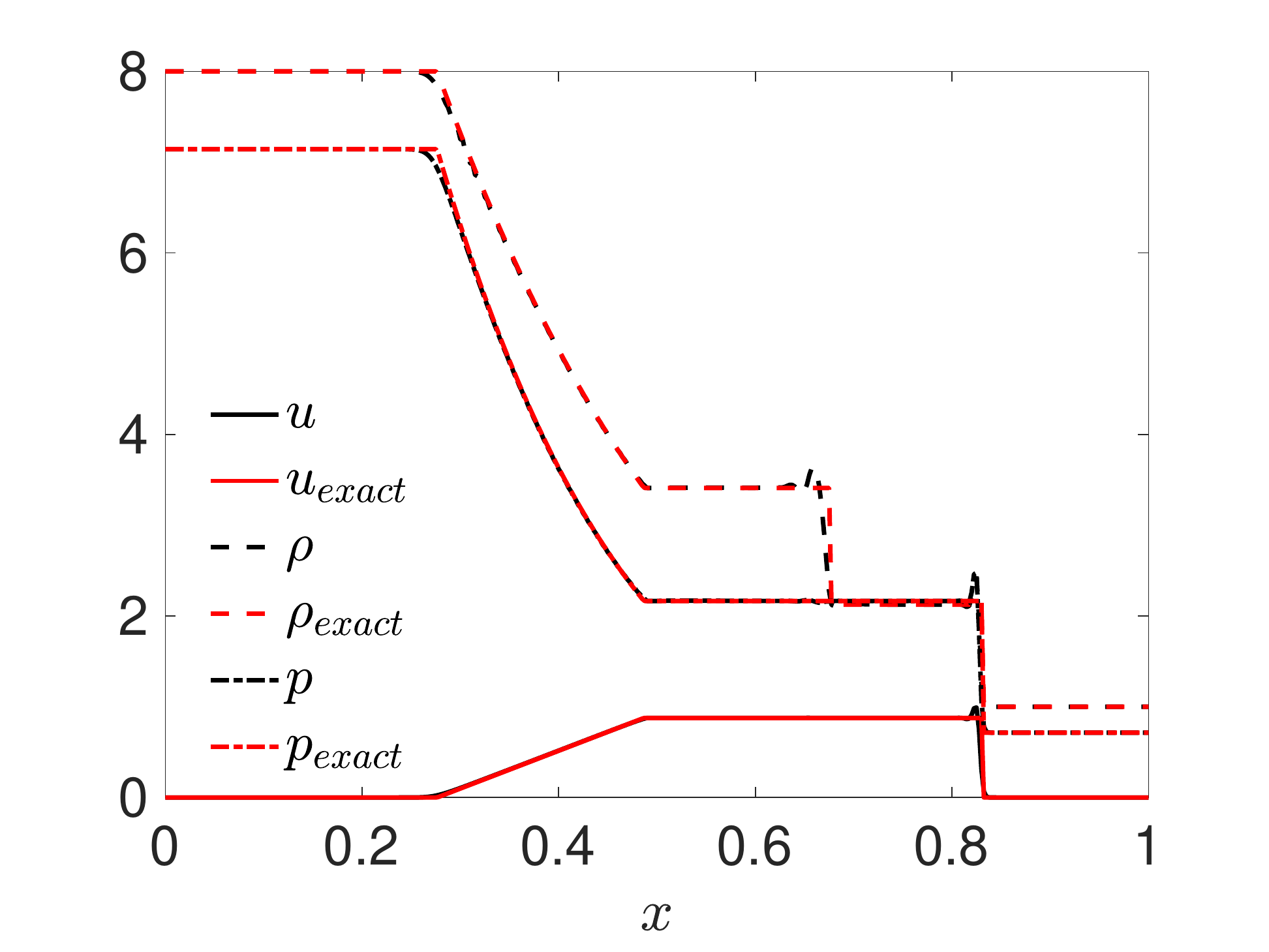}
  \caption{}
  \end{subfigure}
         \begin{subfigure}{0.49 \textwidth}
\includegraphics[width=0.95\textwidth]{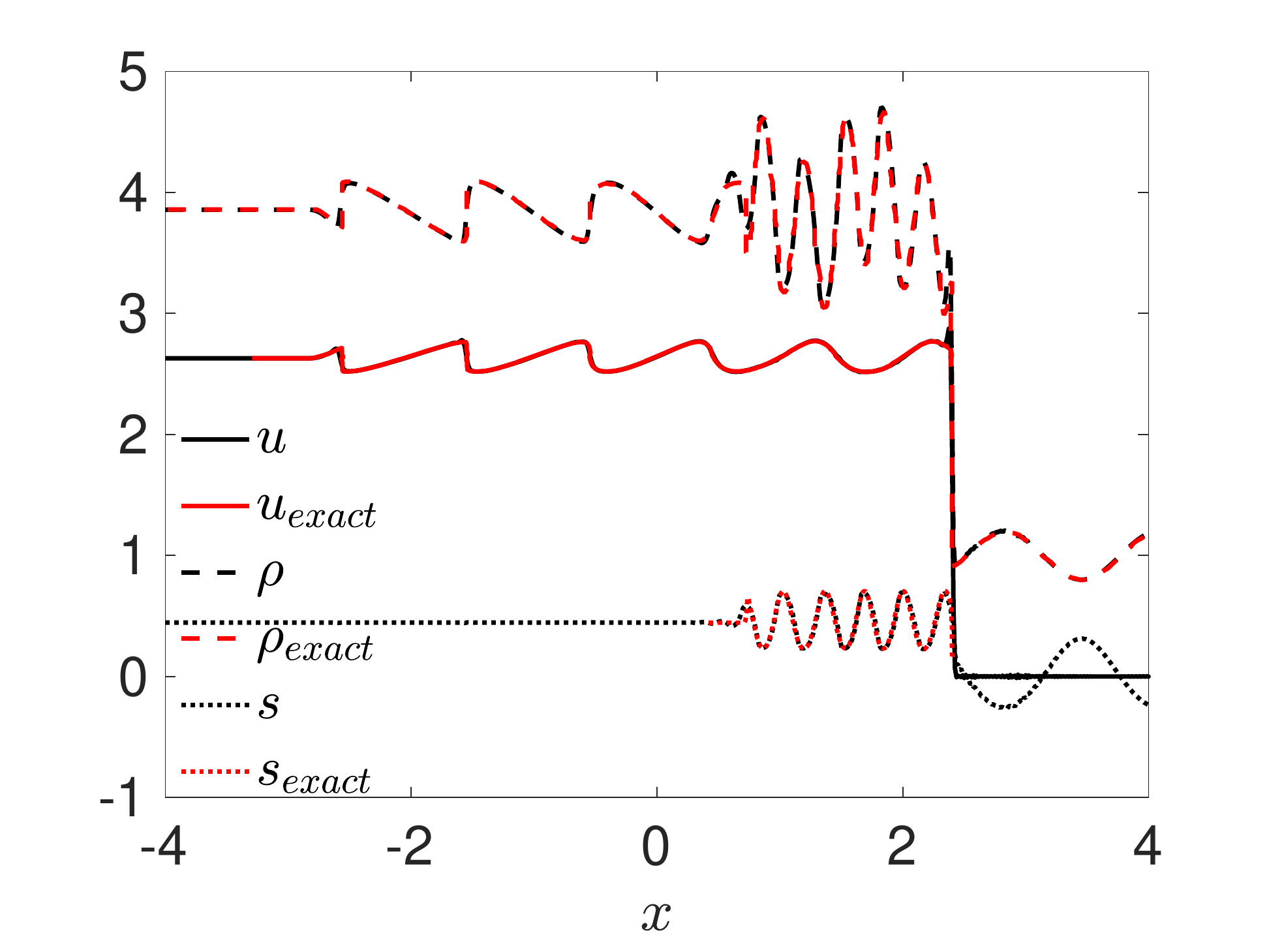}
  \caption{}
  \end{subfigure}
    \caption{Results of simulations using the limiter-inspired LAD model (Eq.~\eqref{eqn:TVD-inspired_D}) plotted for (a) Sod's shock tube and (b) Shu-Osher. Comparison of normalized artificial bulk viscosity values between traditional and limiter-inspired LAD models plotted for (c) Sod's shock tube and (d) Shu-Osher. Mesh-refined (factor of 2) results for the limiter-inspired LAD model in (a, b) plotted for (e) Sod's shock tube and (f) Shu-Osher.}
      \label{fig:TVD-inspired_SO_ST}
\end{figure}

We now test the proposed limiter-inspired LAD model in Eqs.~\eqref{eqn:TVD-inspired_D}--\eqref{eqn:TVD-inspired_D_clipped_filtered} on the canonical problems investigated in Section \ref{subsubsec:results_1D_trad_LAD}, to demonstrate how it can address the aforementioned shortcomings of traditional LAD. First, we consider Sod's shock tube and 1D Shu-Osher. We apply Eq.~\eqref{eqn:TVD-inspired_D} for the mass equation, and Eq.~\eqref{eqn:TVD-inspired_D_clipped_filtered} for the momentum and total energy equations. Figure \ref{fig:TVD-inspired_SO_ST}(a) shows the result for Sod's shock tube with $\Delta x = 0.005$ and $\Delta t=0.0005$, while Figure \ref{fig:TVD-inspired_SO_ST}(b) shows the result of the 1D Shu-Osher test with $\Delta x = 0.025$ and $\Delta t=0.001$. Both of these results are significantly more accurate than the results obtained using the traditional LAD models shown in Figure \ref{fig:LAD_optimal_stuff}. It is useful to compare the normalized values of the artificial diffusivities between the two methods. Figure \ref{fig:TVD-inspired_SO_ST}(c, d) compares the normalized artificial bulk viscosity from the limiter-inspired model (Figure \ref{fig:TVD-inspired_SO_ST}(a, b)) with the traditional LAD model (Figure \ref{fig:LAD_optimal_stuff}(a, b)) for Sod's shock tube and 1D Shu-Osher, respectively. It is interesting that compared to the traditional LAD models, the limiter-inspired LAD is more active in smooth regions close to the discontinuities for both test cases. This is presumably required to dampen the dispersion errors due to CD that would be otherwise observed in such regions. Figure \ref{fig:TVD-inspired_SO_ST}(e, f) refines the mesh for the results in Figure \ref{fig:TVD-inspired_SO_ST}(a, b) by a factor of 2 to demonstrate convergence of the limiter-inspired model's results to the exact solutions.

\begin{figure}
    \centering
    \begin{subfigure}{0.49 \linewidth}
        \includegraphics[width=0.95\textwidth]{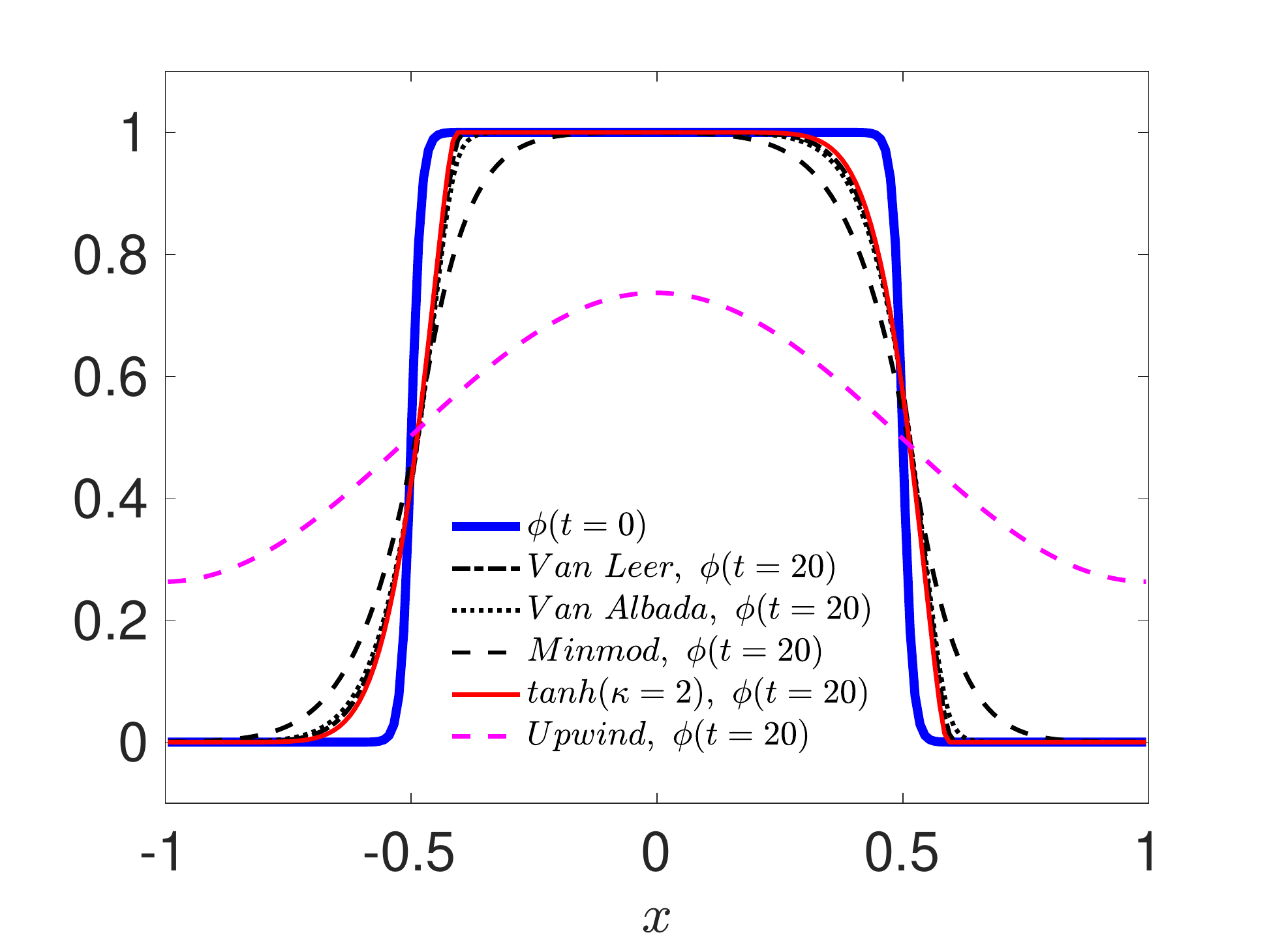}
          \caption{}
          \label{fig:init_drop_advection_mom_cons}
    \end{subfigure}
    \begin{subfigure}{0.49 \linewidth}
        \includegraphics[width=0.95\textwidth]{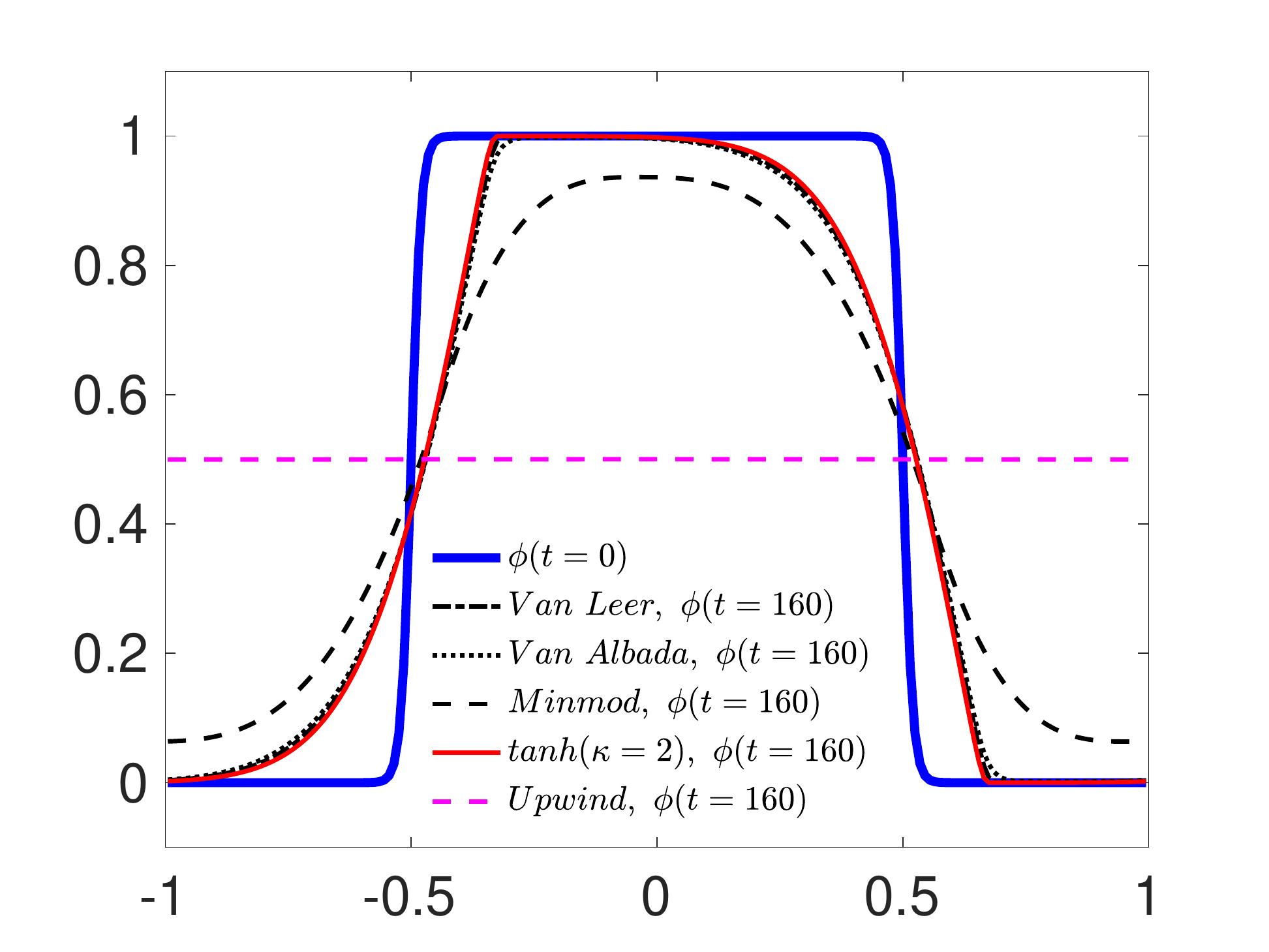}
          \caption{}
          \label{fig:final_inconsist_drop_advection_mom_cons}
    \end{subfigure}
    \caption{Results from scalar advection with Eq.~\eqref{eqn:PDE_pure} using TVD flux limiters and the limiter-inspired LAD model [Eq.~\eqref{eqn:TVD-inspired_D}] plotted at (a) $t_f=20$ and (b) $t_f=160$.}
      \label{fig:1D_scalar_advection_TVD}
\end{figure}

Next, we test the limiter-inspired model on the scalar advection problem introduced in Section \ref{subsubsec:results_1D_trad_LAD}. The results of advection at times $t=20$ and $t=160$ are shown in Figure \ref{fig:1D_scalar_advection_TVD}(a, b). We observe that, at both times, the limiter-inspired LAD model with second-order CD performs better than van Leer, van Albada, minmod, and first-order UD in retaining the profile of $\phi$ during advection. Moreover, the improvement of the proposed model in retaining the $\phi$ profile is significant compared to the traditional LAD method (shown in Figure \ref{fig:1D_scalar_advection_LAD}). Note, finally, the bounded solutions for this advection problem, a property that can be important for robust calculations.

\subsubsection{Extension of limiter-inspired LAD to multiple dimensions}
\label{subsec:multiD_TVD_insp_LAD}

In the limiter-inspired LAD model, the derivative of the flux, rather than the derivative of the primitive variable itself, is used to compute the artificial flux, as illustrated by the simple example in Eq.~\eqref{eqn:TVD_PDE_Model}. As such, the multi-dimensional extension of the algorithm becomes nontrivial. As explained in Section \ref{subsec:TVD_LAD}, solving Eq.~\eqref{eqn:PDE_pure} using TVD flux limiters for evaluation of advective face fluxes is equivalent to using CD to discretize Eq.~\eqref{eqn:TVD_PDE_Model} with limiter-based diffusivity, as illustrated by the simple example in Eq.~\eqref{eqn:VanLeer_D}. Consider now the generalization of Eq.~\eqref{eqn:PDE_pure} to multiple dimensions,
\begin{equation}
    \frac{\partial \phi}{\partial t}+\nabla\cdot \vec{f}=0.
    \label{eqn:PDE_pure_multiD}
\end{equation}
Instead of using TVD flux limiters to obtain the advective flux ($\vec{f}=\vec{u}\phi$) on the cell faces, we may discretize the multi-dimensional extension of Eq.~\eqref{eqn:TVD_PDE_Model} using central spatial operators. However, we need to take into account the topology of the numerical grid. For example, on a 2D Cartesian grid we can use CD to discretize:
\begin{equation}
    \frac{\partial \phi}{\partial t}+\nabla\cdot \vec{f}=\frac{\partial}{\partial x}\left(D_x^\ast\frac{\partial f_x}{\partial x}\right)+\frac{\partial}{\partial y}\left(D_y^\ast\frac{\partial f_y}{\partial y}\right).
    \label{eqn:TVD_PDE_Model_multiD}
\end{equation}
For the limiter-based model, calculation of $D_x^\ast$ involves $\Delta_x$, $\text{sgn}(u_x)$, $f_x$, and their derivatives in the $x-$direction, for example using Eq.~\eqref{eqn:VanLeer_D} for the van Leer limiter and a uniform grid cell size; the calculation of $D_y^\ast$ proceeds in a similar fashion. Alternatively, one can compute $D_x$ and $D_y$ for Eq.~\eqref{eqn:TVD_PDE_Model_multiD} via the limiter-inspired model given by Eq.~\eqref{eqn:TVD-inspired_D}--\eqref{eqn:TVD-inspired_D_clipped_filtered}. For instance, for AMD, a limiter-inspired $D_x^\ast$ is computed via CD evaluation of:
\begin{equation}
    D_x^\ast=\frac{\Delta x}{2}\text{sgn}(u_x)\text{tanh}\left[2\frac{\frac{\partial f_x}{\partial x}-\frac{\partial \overline{f_x}}{\partial x}+\frac{\Delta x}{2}\text{sgn}(u_x)\frac{\partial^2 f_x}{\partial x^2}}{\kappa\frac{\partial {f_x}}{\partial x}}\right],
    \label{eqn:TVD-inspired_multiD_x}
\end{equation}
on the $x-$faces, where filtering involves only neighbors in the $x$ direction, and we have again assumed a uniform grid cell size. Similarly, $D_y^\ast$ is computed via CD evaluation of:
\begin{equation}
    D_y^\ast=\frac{\Delta y}{2}\text{sgn}(u_y)\text{tanh}\left[2\frac{\frac{\partial f_y}{\partial y}-\frac{\partial \overline{f_y}}{\partial y}+\frac{\Delta y}{2}\text{sgn}(u_y)\frac{\partial^2 f_y}{\partial y^2}}{\kappa\frac{\partial {f_y}}{\partial y}}\right],
    \label{eqn:TVD-inspired_multiD_y}
\end{equation}
on the $y-$faces, where filtering involves only neighbors in the $y$ direction.

Next, we demonstrate the application of the limiter-based LAD method introduced in Eqs.~\eqref{eqn:TVD_PDE_Model_multiD}--\eqref{eqn:TVD-inspired_multiD_y} to the two-dimensional (2D) version of the scalar advection problem discussed in Sections \ref{subsubsec:results_1D_trad_LAD} and \ref{sec:results_tvd_inspired}. Clearly, the above methodology is limited to Cartesian grids, and, more importantly, \emph{lacks rotational invariance}. We defer a rotationally-invariant formulation, which is applicable to both structured and unstructured grids, to future work.

\subsubsection{2D scalar advection problem}
\label{sec:results_2d_tvd_inspired}
\begin{figure}
    \centering
    \begin{subfigure}{0.49 \linewidth}
        \includegraphics[width=0.95\textwidth]{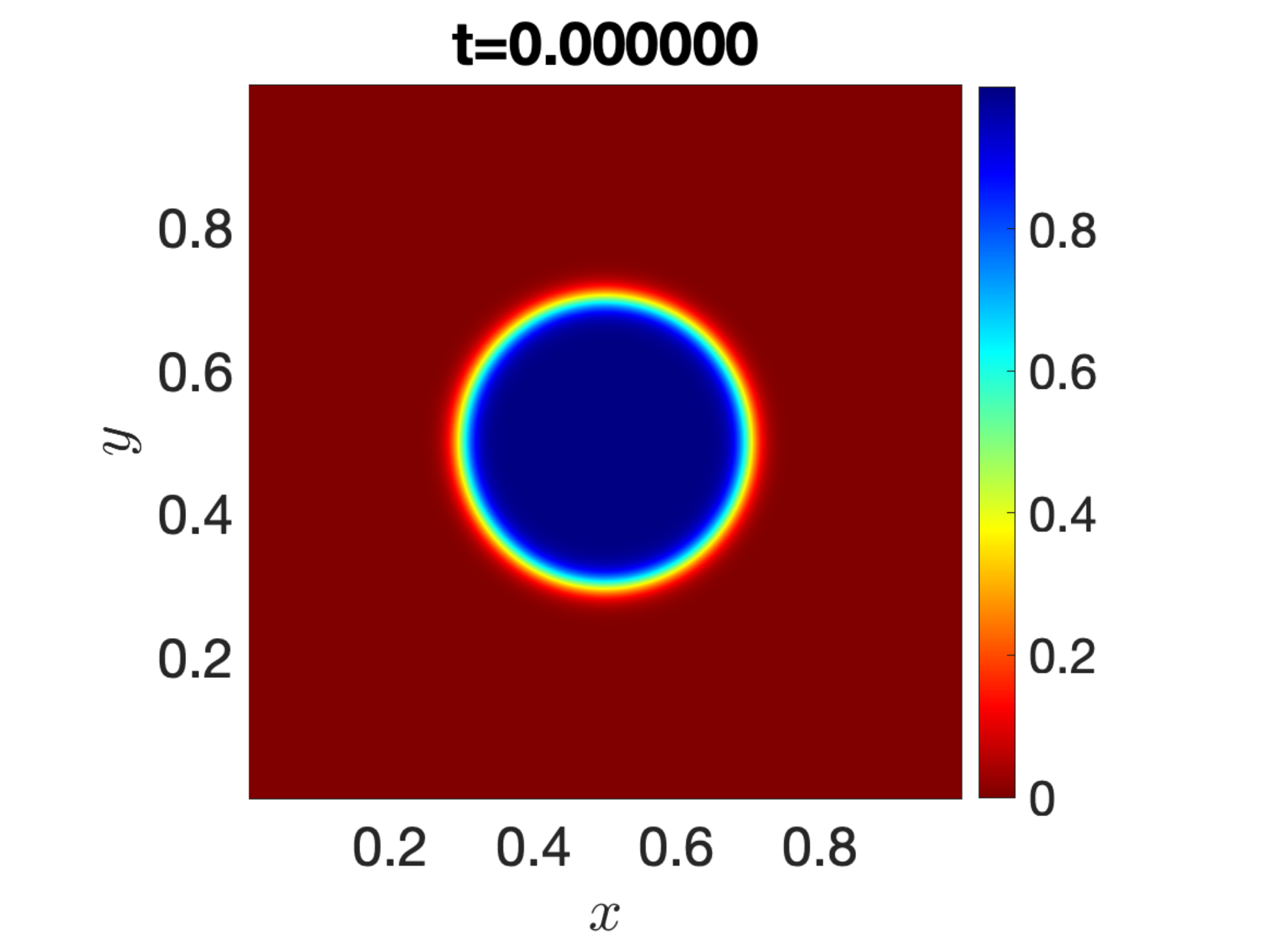}
          \caption{}
          \label{fig:init_drop_2D_advection}
    \end{subfigure}
    \begin{subfigure}{0.49 \linewidth}
        \includegraphics[width=0.95\textwidth]{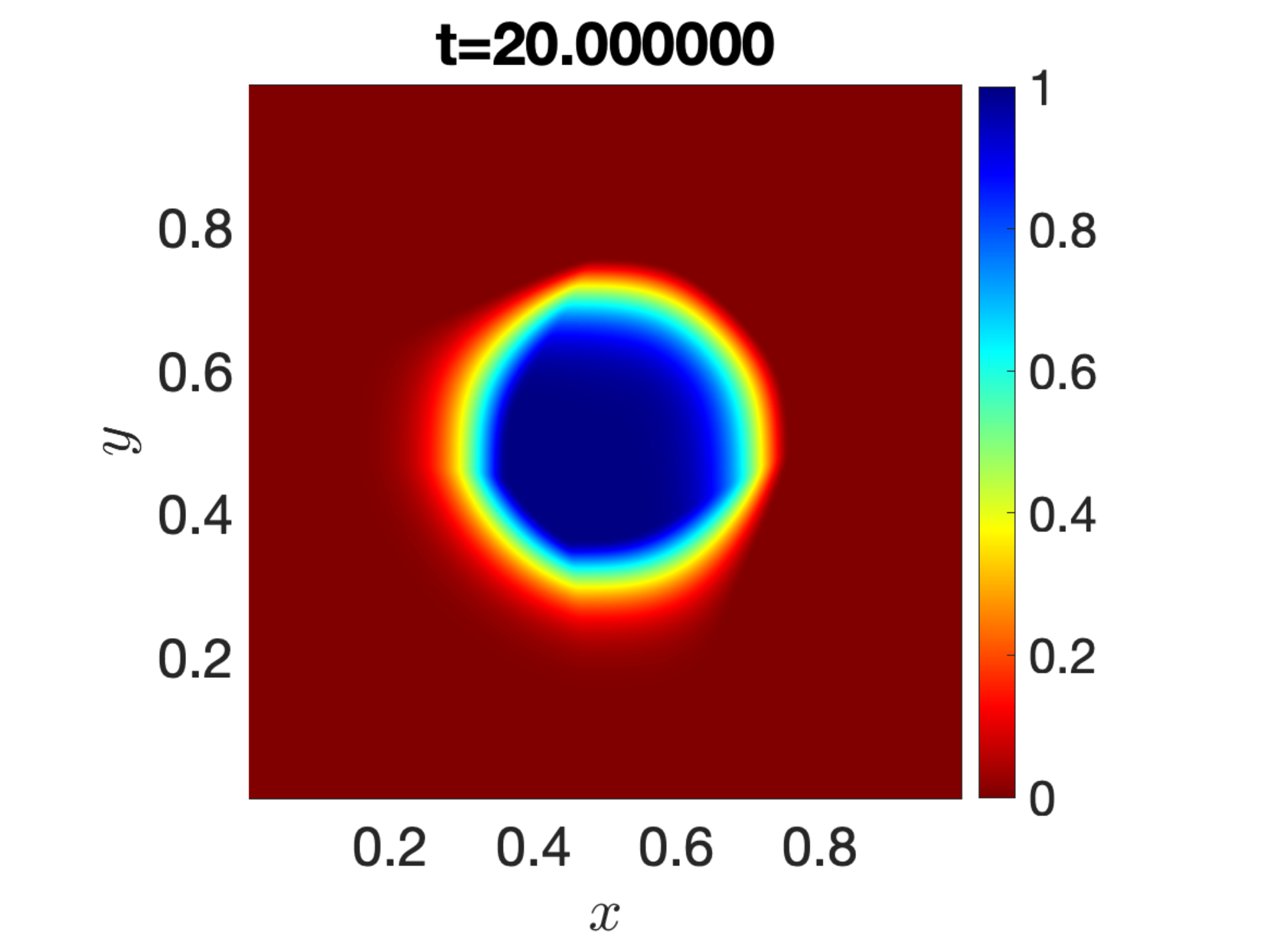}
          \caption{}
          \label{fig:2D_advection_20}
    \end{subfigure}
       \begin{subfigure}{0.49 \linewidth}
        \includegraphics[width=0.95\textwidth]{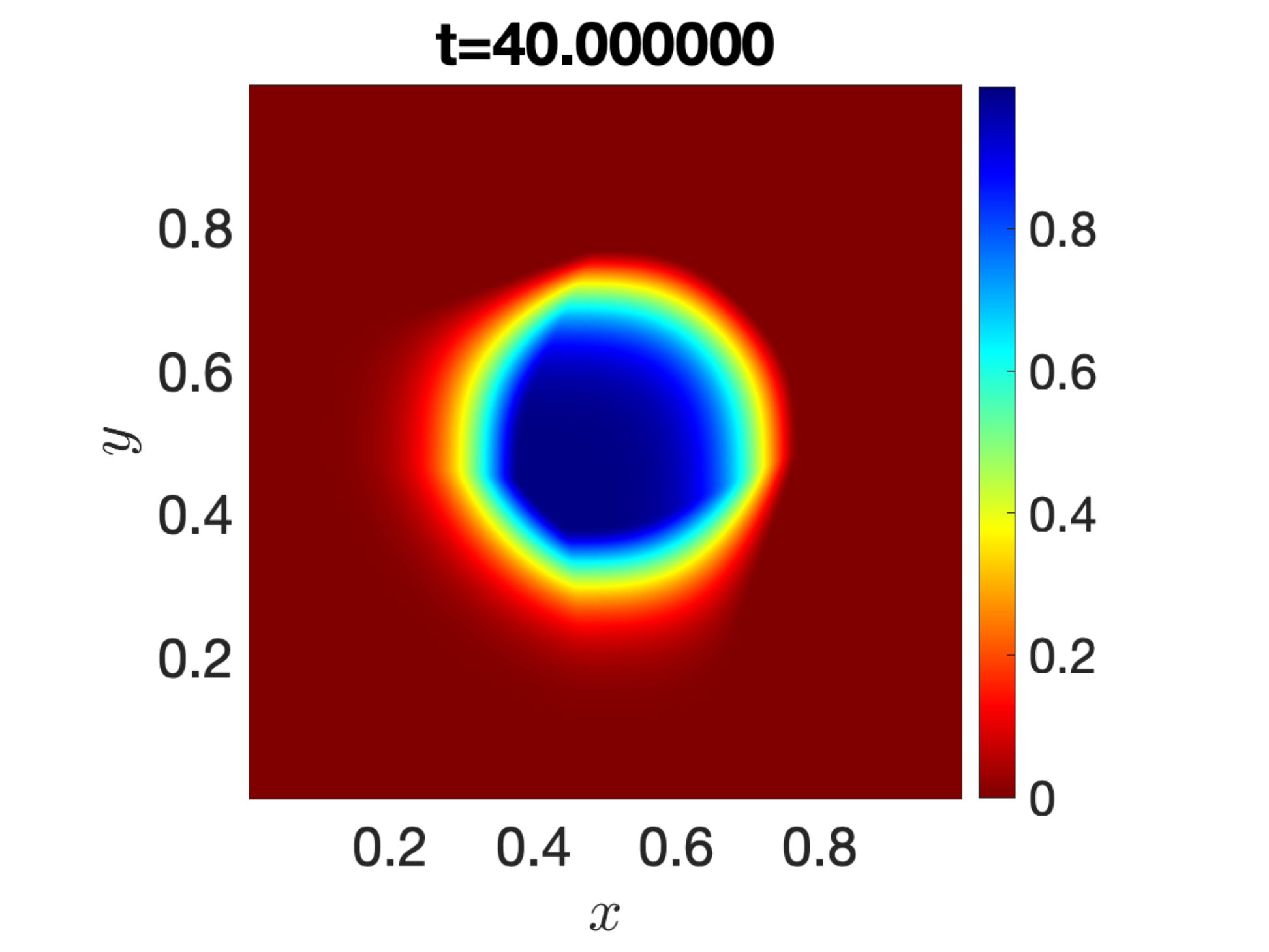}
          \caption{}
          \label{fig:2D_advection_40}
    \end{subfigure}
    \begin{subfigure}{0.49 \linewidth}
        \includegraphics[width=0.95\textwidth]{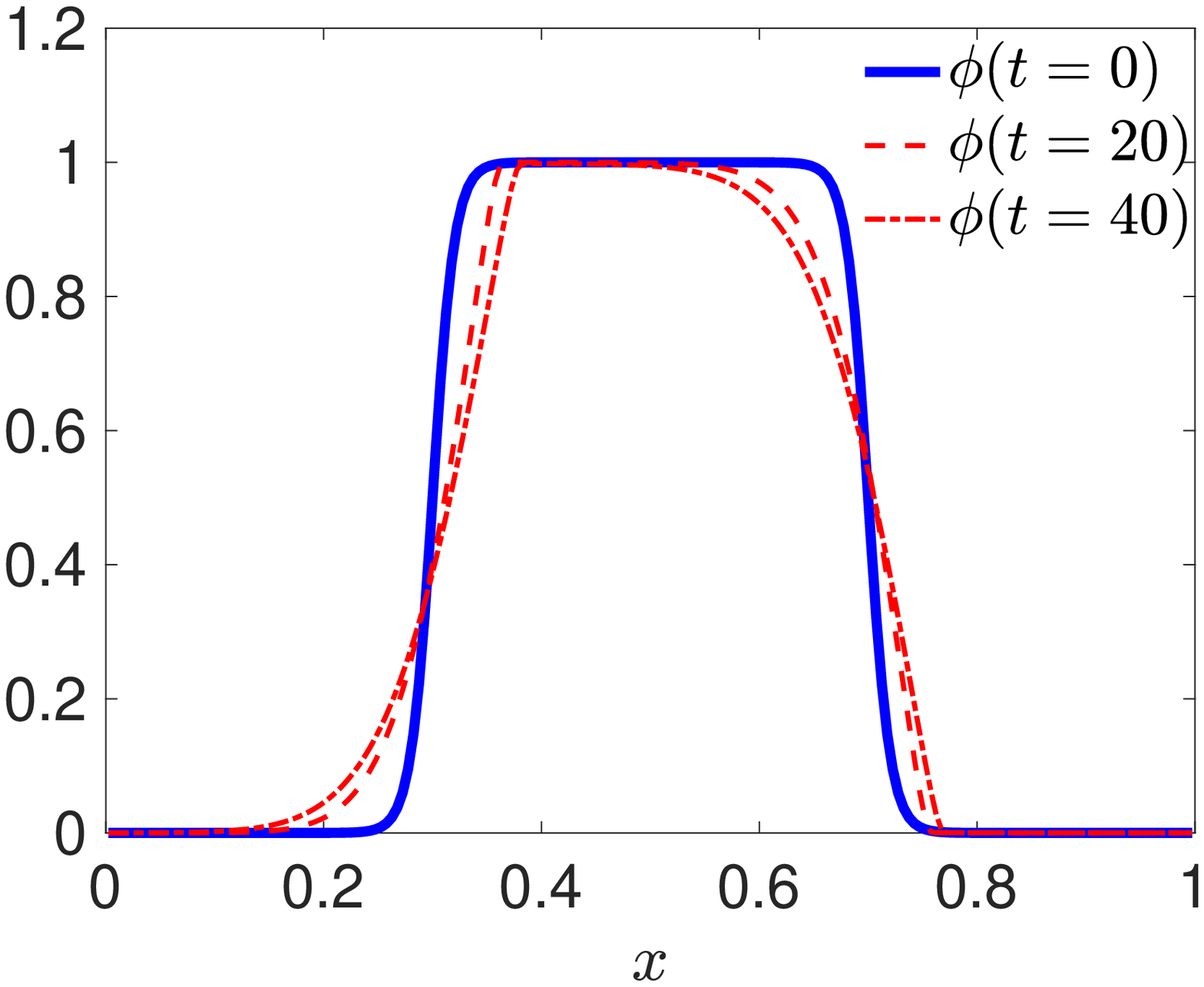}
          \caption{}
          \label{fig:2D_cross_section}
    \end{subfigure}
    \caption{Results from scalar advection governed by Eq.~\eqref{eqn:PDE_pure_multiD} using the multi-dimensional limiter-inspired LAD model [Eqs.~\eqref{eqn:TVD_PDE_Model_multiD}--\eqref{eqn:TVD-inspired_multiD_y}] plotted at (a) $t=0$, (b) $t=20$, (c) $t=40$, and (d) the solutions as a function of $x$ on the $y=0.5$ line.}
      \label{fig:2D_scalar_advection_TVD}
\end{figure}

We test the multi-dimensional version of the limiter-inspired model, presented in Section \ref{subsec:multiD_TVD_insp_LAD}, in a 2D setting. For this purpose, we consider a scalar advection problem, similar to the tests in Sections \ref{subsubsec:results_1D_trad_LAD} and \ref{sec:results_tvd_inspired}. Specifically, in a nondimensional $1\times1$ domain discretized with a $200\times200$ mesh, we initialize a disk of radius $0.2$ at $(0.5,0.5)$, and advect it with velocity $\vec{u}=(1,1)$. The initial condition is shown in Figure \ref{fig:init_drop_2D_advection}. The results at times $t=20$ and $t=40$ are shown in Figures \ref{fig:2D_scalar_advection_TVD}(b, c). The $y=0.5$ cross-sections of the solutions are shown at times $t=0, 20, 40$ in Figure \ref{fig:2D_cross_section}. From these results, it is evident that the multi-dimensional limiter-inspired LAD is successful in damping the oscillations associated with CD, and yields bounded results similar to its 1D counterpart, shown in Figure \ref{fig:1D_scalar_advection_TVD}. Again, bounded solutions are an indication of the robustness of the method.

\section{Generation of directly computable models via IAT-based strategies}
\label{sec:dir_comput_models}

The aim of the DARPA Computable Models (CompMods) program was to spur the development of strategies yielding \textit{directly computable models}, i.e., methods that reduce the level of effort (LoE) to go from mathematical problem formulation to numerical implementation. The success of such strategies was measured via key metrics including \textit{computability} (i.e., the LoE in generating a readily-usable numerical implementation), \textit{composability} (i.e., the LoE for incorporating multiple physical components), \textit{discrete conservation} (i.e., guaranteeing conservation at the level of the numerical implementation), \textit{computational efficiency} (including load balancing and scalability), \textit{numerical accuracy} and \textit{scale range}. Below we evaluate IAT in terms of the first four metrics \footnote{The various performers in CompMods targeted different aspects of the overarching goal of building directly computable models. Our focus was mainly on reduction of LoE for computability and composability, as well as enforcing discrete conservation by consistently modifying the governing PDEs and, to a lesser extent, improving computational efficiency. We did not address the issues of improving the numerical accuracy or increasing the scale range of the computations.}, and provide details on how IAT aims to satisfy each of them.

\subsection{Computability}
Development of state-of-the-art discontinuity-capturing spatial discretization schemes like VoF (for multi-phase flows) and MUSCL/ENO/WENO/TENO (for compressible flows) has taken years, resulting in highly-specialized algorithms. Such methods have been incorporated into custom solvers, like NASA's \vulkan \cite{NASA2021}, or have been ported to popular open-source frameworks like \openfoam, by domain experts intimately familiar with the intricacies of these complex schemes \cite{gartner2020}. In practice, when faced with a rapidly emerging multi-physics problem, this creates a barrier to quick implementation for novice computational scientists who either need to familiarize themselves with the manual of a bespoke framework, or figure out how to integrate and work with custom libraries in open-source software like \openfoam.

Instead, by augmenting the original governing PDEs with artificial terms representing dissipation, heat conductivity, discontinuity sharpening, etc., IAT makes it possible to use non-dissipative spatial schemes like CD in the numerical implementation, with various orders of spatial accuracy. IAT thereby shields the end user from dealing with the complexities associated with minimizing and/or tuning the dissipation of specialized discontinuity-capturing discretization schemes, or other intricacies associated with learning to use custom solvers or libraries. Moreover, IAT facilitates the automation of the conversion from modified PDEs to numerical source codes readable by \openfoam or other open-source frameworks through the use of grid templates.

Our approach for simulating the laminar hypersonic compression corner was a concrete example of IAT's ability to generate directly computable models. Within a span of only six months, we were able to derive appropriate PDE modifications adhering to the requirements of IAT, and generate \openfoam source code that could predict wall-integrated heat flux, pressure and shear stress to within 12\% error compared to the state-of-the-art (SOA) solver \vulkan serving as the ground truth.

\subsection{Composability}

When dealing with emerging multi-physics problems, the ability to easily incorporate additional physics components into the problem formulation and translate them into a numerical implementation is of paramount importance. Specialized numerical algorithms are typically only validated for use under a limited set of applicability conditions, and may break unexpectedly when that underlying set of assumptions is violated. Incorporating additional physical processes may require changing existing governing equations and/or adding new ones, which could invalidate some of these assumptions and require surgical modifications to the existing code to make it work in the new scenario. Such changes may take time, especially when needed to be undertaken by non-experts who are not well-versed in the original source code.

In IAT, designing additional artificial terms either for existing or new governing equations also requires additional LoE along with physical insight and expertise (partly generic and partly domain-specific). However, within the context of rapid-response scenarios, one can envision a parallel workflow with computational scientists performing simulations with off-the-shelf numerics discretizing an initial set of modified PDEs, while the team of applied mathematicians who designed the original PDE modifications is deriving additional artificial terms for a modified/extended set of governing equations accounting for the inclusion of additional physics components into the problem formulation. Moreover, conversion of the new set of modified governing PDEs into numerical source code will not require major changes to the existing multi-physics compiler automating this conversion, as a similar set of operators will be present in the new set of modified PDEs. Hence, we posit that the overall LoE related to composability will be lower for IAT than for SOA bespoke spatial discretization schemes for capturing discontinuities in fluid flows.

\subsection{Discrete conservation}

Discontinuity-capturing modifications, both at the PDE-level and in numerical methods/code, have to be {\it consistent} across coupled equations to ensure that non-negotiable first principles (e.g., conservation laws) remain intact. For instance, for multi-phase flows, it is well-known that artificial diffusion and sharpening of the phase interfaces in the conservation of mass equation should consistently be accounted for in the conservation of momentum equation to make sure that conservation of kinetic energy follows automatically \citep{Huang2021,Mirjalili2021,Terashima2013,Haga2019,Jain2020}. Enforcing consistency via equation modification is typically straightforward, as it leverages analytical proofs. Additionally, the use of standard central operators allows for analytical properties like conservation to extend to the discrete realm. This is in contrast to enforcing consistency for complex schemes like VoF, where complex geometrical algorithms must be considered \cite{Mirjalili2021}.

\subsection{Computational efficiency}

As IAT captures discontinuities at the PDE level by augmenting the governing equations with artificial terms, it enables the use of low-cost numerical recipes like central differencing everywhere in the domain, i.e., the discretization strategy used with IAT is agnostic to the location of discontinuities. Compared to the deployment of complex numerics, IAT can thus result in considerable computational savings \footnote{Discretization of the additional PDE terms introduced by IAT using non-dissipative schemes like central differencing results in a negligible computational overhead during subsequent simulations. Furthermore, while certain terms are only ``activated'' near the discontinuities by virtue of the numerical values of the parameters they depend on, no logical operations (e.g., ``if-then-else'' statements) are required for detecting discontinuities. This results in a balanced distribution of the computational effort throughout the domain, making load-balancing straightforward \citep{Mirjalili_comparison,Jain2020}.}

In two-phase flows, in contrast to phase-field methods, VoF schemes rely on sensors (related to the volume fraction) to detect phase interfaces, and apply complex, often expensive, algebraic and/or geometric operations in the presence of such interfaces. In such cases, the additional computational cost in regions surrounding discontinuities makes load-balancing a non-trivial task \citep{Jofre2015,Mirjalili_comparison}.

For compressible flows, popular shock-capturing schemes like MUSCL or WENO/TENO have parameters that adapt in the vicinity of shocks/contact discontinuities, making the computational cost uniform, and the computations scalable. However, the calculation of smoothness indicators and the common use of characteristic projections render these schemes more expensive than standard central schemes. As a result, some researchers have resorted to hybrid approaches that combine these schemes with lower-cost, lower-dissipation spatial schemes that are used away from discontinuities \citep{Zhao2019,Pirozzoli2002}. These modifications introduce logical statements into the numerical implementation and harm the scalability of the overall model.

\section{General strategies for PDE modification}
\label{sec:PDE_mod_strategies}

Section \ref{sec:IAT_reqs} listed the requirements that a given PDE modification must satisfy to be considered IAT. Proposing PDE modifications, however, is a non-trivial, non-routine task that requires expertise, and sometimes, intuition. Within this work, we have introduced several instances of IAT. The corresponding PDE modifications were obtained using the following strategies for capturing discontinuities in fluid flows:

\begin{enumerate}
    \item \textbf{Taking inspiration from physical phenomena and equations rooted in the thermodynamics of the discontinuity to derive phenomenological models.} The conservative phase-field model of Eq.~\eqref{eqn:2phase_pf} (see Section \ref{subsec:modifiedPDE_verification_IAT_req}) was derived using this approach by manipulating the Allen--Cahn equation \citep{Chiu_and_Lin} rooted in Van der Waals' equation of state. Similarly, artificial diffusivity models in compressible flows were originally developed by taking inspiration from physical diffusion \citep{Cook2007}.
    \item \textbf{Using analogies, consistency checks (including thermodynamic consistency), and physical intuition to postulate the form of PDE modifications.} This approach is also quite common and is especially useful in systems of coupled equations. An example is \cite{Mirjalili2021}, wherein conservation of kinetic energy, consistency of reduction, and the dual meaning of velocity were used to derive the correction to the momentum transport equation for mass-momentum consistency. Another example is the approach used in \cite{Jain2019} to derive the model for confined scalar transport in two-phase flows using phase-field methods.
    \item \textbf{Assuming a reasonable microstructure for the discontinuity to derive PDE modifications for new physical effects.} This approach would typically rely on analytical treatments and homogenization to derive the functional dependence of the PDE modifications on variables varying across the discontinuity. This approach was leveraged in \cite{Mirjalili2022} to derive heat/mass transfer terms when using phase-field models for simulation of two-phase systems.
    \item \textbf{Taking inspiration from numerical methods that have desirable properties (e.g., the TVD property, boundedness, and localization) to reverse engineer PDE terms.} We use this approach in Section \ref{subsec:TVD_LAD} to derive localized artificial diffusivity terms. Note that after this conversion, the user is free to use standard numerical methods while still capturing discontinuities robustly and accurately.
\end{enumerate}

While other strategies may be envisioned for designing PDE modifications, the above options can be considered powerful and versatile tools in the toolbox of applied physicists interested in the IAT approach.

\section{Conclusions and outlook}
\label{sec:conclusions}

We introduced inverse asymptotic treatment (IAT) as a strategy for enabling {\it directly computable} modeling of fluid flows with dynamically evolving discontinuities. This approach modifies the partial differential equations (PDEs) governing such flows, making it possible to use standard numerical schemes like central difference (CD) spatial operators to capture these discontinuities. This facilitates automating the conversion of the modified PDEs into a directly computable numerical implementation that can be run on readily-available computational frameworks, e.g., by means of a ``multi-physics'' compiler that uses grid templates.

After listing the requirements for IAT, we explained this strategy in the context of two-phase flows, highlighting its ability to systematically enforce consistency. Next, focusing on single-phase compressible flows, our research led to the following major conclusions:
\begin{enumerate}
    \item Using localized artificial diffusivities (LADs) can be viewed as an instantiation of IAT, whereby shock and contact discontinuities are artificially thickened to avoid spurious oscillations resulting from the use of non-dissipative numerical schemes such as second-order CD, while yielding a numerical solution that converges to the exact solution as the artificial thickness is reduced to zero.
    \item Through numerical tests, we showed that using second-order CD combined with traditional LAD techniques can lead to a trade-off between dispersion errors (undershoots/overshoots) and numerical dissipation, as a result of which, optimizing the free parameters becomes a difficult, problem-dependent task. Nevertheless, we showed that for the case of a 2D laminar hypersonic compression corner, traditional LAD yielded comparable results to NASA’s \vulkan for small corner angles, and predicted all wall-integrated metrics with a relative error of less than 12\% for all corner angles considered. However, a clear discrepancy remained in the prediction of the flow separation and reattachment locations, and it remains an open question whether further optimization of the free parameters could lead to improved accuracy.
    \item We proposed a new tuning-free LAD method inspired by flux limiters endowing the total variation diminishing (TVD) property to certain high-resolution numerical schemes. Compared to traditional LAD, this ``limiter-inspired'' LAD displayed superior performance on a 1D scalar advection problem as well as on the commonly used benchmarks of Sod’s shock tube and 1D Shu-Osher. Finally, we discussed the multi-dimensional extension of this limiter-inspired approach for the case of Cartesian grids, and demonstrated its performance on the 2D scalar advection of a disk.
\end{enumerate}

The IAT strategy can be particularly appealing to new users who are not necessarily well-versed in domain-specific numerical schemes. IAT still requires insight (partly generic and partly domain-specific) to appropriately modify the governing PDEs; however, once such modifications are done by domain experts, standard numerical schemes (e.g., second-order CD) can be used safely by domain experts and non-experts alike in their numerical platform of choice. This shift from specialized to standard numerics, albeit at a nontrivial cost to design appropriate modifications to the governing PDEs, can democratize computation by facilitating the use of standard numerical implementations by a wider range of computational scientists, thereby reducing the level of effort/time needed to go from ideation and problem formulation to numerical simulation. This is especially valuable when a quick response to rapidly emerging problems is required, such as rapid evaluation of novel hypersonic vehicle designs. In those situations, IAT can aid in quickly iterating numerical solvers (including higher-order formulations) suitable for a range of interacting physics scenarios and corresponding design requirements.

\section*{Acknowledgments}
This material is based upon work supported by the Defense Advanced Research Projects Agency (DARPA) under agreement no. HR00112090065 (DARPA Computable Models program).

S. Mirjalili would like to thank Jacob West for providing helpful comments on an earlier version of this work \citep{Mirjalili_TVD}, particularly the suggestion to clip and filter the TVD-inspired diffusivities when possible.

Furthermore, the authors would like to thank Matthew D. O'Connell, Nathaniel Hildebrand, and Meelan Choudhari from NASA, who served as the independent validation and verification (IV\&V) team for the DARPA Computable Models program, for providing us with the compression corner challenge problem and offering invaluable guidance throughout the program to refine IAT. The authors want to give special thanks to Matthew D. O'Connell for providing them with the NASA \vulkan results for comparison purposes.

\bibliographystyle{unsrtnat}

\end{document}